\documentclass{aa}  
\usepackage{graphicx}
\usepackage{graphicx,subcaption} 
\usepackage{natbib}
\usepackage{txfonts}

\usepackage{hyperref}

\usepackage[font=small,labelfont=bf]{caption}

\usepackage{newtxtext,newtxmath}
\usepackage[T1]{fontenc}
\usepackage{amsmath}	
\usepackage{booktabs}
\usepackage{multirow}

\titlerunning{}
\authorrunning{Renu et al.}

\begin{document} 
    
\title{Investigating the role of bars in quenching star formation using spatially resolved UV-optical colour maps}

   \author{D. Renu
          \inst{1,2}, 
          Smitha Subramanian\inst{1},
          Suhasini Rao\inst{1,3},
           \and
          Koshy George\inst{4}
          }
   \institute{Indian Institute of Astrophysics, 2nd Block Koramangala, Bengaluru, 560034\\
             \email{renukadyan3232@gmail.com, renu.devi@iiap.res.in} 
         \and
            Pondicherry University, R.V. Nagar, Kalapet, 605014, Puducherry, India
             \and
             Department of Physics, University of Alberta, CCIS 4-181, Edmonton, AB T6G 2E1, Canada
              \and
             Faculty of Physics, Ludwig-Maximilians-Universitat, Scheinerstr. 1, Munich 81679, Germany
             }

   \date{Received XXXX ; accepted XXXX}
 
  \abstract
   {\textit{Context}: Bars are ubiquitously found in disc galaxies and are known to drive galaxy evolution through secular processes. However, the specific contribution of the bars in the suppression of star formation is still under inspection. \\
   \textit{Aims}: Our aim is to investigate the role of bars in quenching star formation using spatially resolved UV-optical colour maps and radial colour profiles of a sample of 17 centrally quenched barred galaxies in the redshift range of 0.02 -- 0.06. \\
\textit{Methodology}: The sample of centrally quenched barred galaxies are selected based on their location in the M$_{\star}$ - SFR plane. They are classified as passive based on the parameters from the Max Planck Institute for Astrophysics (MPA) and Johns Hopkins University (JHU) value-added catalogue (MPA - JHU VAC), but are classified as non-passive based on the parameters from the GALEX-SDSS-WISE Legacy (GSWLC) catalogue, indicating a passive inner region and recent star formation in their extended disc.  We use the archival SDSS optical r-band and GALEX far- and near-UV imaging data of these galaxies and created spatially resolved (FUV$-$NUV vs NUV$-$r) colour-colour maps to understand the nature of the UV emission from different regions of these galaxies. We also analyse their NUV$-$r radial profiles and use the NUV$-$r colour as a proxy for the stellar population age in the different regions of these galaxies. 
A control sample of 8 centrally quenched unbarred galaxies is also analysed to disentangle the effect of bulge and bar in quenching star formation. \\
\textit{Results}: The centrally quenched barred galaxies display redder colours (NUV$-$r $>$ 4 --  4.5 mag) in the inner regions,  up to the length of the bar, indicating the age of stellar population in these regions to be older than $>$1 Gyr. Most barred galaxies in our sample host pseudo bulges and do not host AGN, indicating that the most probable reason for the internal quenching of these galaxies is the action of stellar bar. In comparison to their unbarred counterparts, lying in a similar regime of stellar mass and redshifts, the barred galaxies show redder colours (NUV$-$r $>$ 4 mag) to a larger spatial extent.\\
\textit{Conclusions}: Bars in their later stages of evolution turn the inner regions of galaxies redder, leading to quenching, with the effect being most prominent up to the ends of the bar and creating a region dominated by older stellar population. This may occur because bars have already funneled gas to the galactic centre leaving behind no fuel for further star formation. Spatially resolved studies of a larger sample of barred galaxies, at different redshfits, will provide more insights to the role of bar in quenching star formation and different evolutionary stages of quenching.}

   \maketitle

%

\section{Introduction}
\label{sec:intro}
The suppression of star formation in galaxies is known as quenching of star formation. It plays an important role in the evolution of galaxies \citep{man2018star}. The connection of quenching to galaxy evolution is observed in the optical colour distribution of galaxies in the local Universe. It shows a bimodal distribution, with blue region mostly populated by star-forming galaxies and the red region dominated by galaxies with little or no ongoing star formation (known as passive galaxies; \citealt{strateva2001color, baldry2004quantifying}). Such a bimodality in the rest frame colour distribution of galaxies exists up to a redshift of z $\sim$ 1. However, from z $\sim$ 1 to the present epoch, the population of blue, luminous galaxies is gradually declining, while the number of red galaxies is consistently growing.
 This implies that a good fraction of blue galaxies stop star formation and become part of the red sequence \citep{bell2004nearly, faber2007galaxy}. 
Another manifestation of this bimodal optical colour distribution can be seen in the star formation rate (SFR) vs stellar mass (M$_{\star}$) plot \citep{salim2007uv}. The three-dimensional SFR - M$_{\star}$ relation of local galaxies (with the third dimension accounting for the number of galaxies in the SFR - M$_{\ast}$ bins) in the Sloan Digital Sky Survey (SDSS) database show two prominent peaks, one for star-forming galaxies and one for the quenched ones \citep{renzini2015objective}. The star-forming galaxies show a clear correlation with M$_{\star}$ and form a main sequence (MS) in the two-dimensional SFR - M$_{\star}$ plane. The position of galaxies from the MS is widely used to identify galaxies that do not belong to the MS, such as starburst outliers on one side and, the quenched/passive galaxies on the other side (\citealt{2011ApJ...742...96W, 2017A&A...597A..97M, Guo_2019}).\\
Identifying the physical mechanism responsible for star formation quenching in galaxies is one of the open questions. Both internal (feedback from the active galactic nuclei (AGN) and stars, action of stellar bars, bulges) as well as external (ram pressure stripping, major mergers, harassment, starvation, strangulation) mechanisms are proposed to be responsible for quenching (see \citealt{peng2015, man2018star} and references therein). For external processes like ram pressure stripping or strangulation, star formation quenching is expected to happen outside-in, while internal processes like action of bulge/bar/AGN would show an inside-out trend of quenching \citep{2019ApJ...872...50L}. \cite{george2021bar} suggested that different SFR indicators can be used to identify centrally quenched galaxies, those which are undergoing inside-out quenching, in the local Universe. In this study, we aim to focus on the barred spiral galaxies which are centrally quenched with extended star-forming discs, to mainly understand the effect of bars in quenching star formation (\citealt{gavazzi2015role, spinoso2016bar, khoperskov2018bar, george2019insights}). \\
Bars are known to be non-axisymmetric structure in the centre of spiral galaxies that can survive for longer time scales (in Gyrs), even though it only takes million years to form a bar \citep{1993RPPh...56..173S, Kraljic_2012, 10.1093/mnras/staa1104}. \cite{2013ApJ...779..162C} found that the likelihood of a galaxy to host a bar is anti-correlated with the specific SFR, regardless of stellar mass or prominence of bulge. Barred galaxies are also found to have lower star formation activity compared to unbarred galaxies (\citealt{2017A&A...598A.114C,Kim_2017}). \cite{2020MNRAS.499.1116F} analysed the stellar populations and star formation histories of 245 barred galaxies from the Mapping Nearby Galaxies at APO (MaNGA) galaxy survey, and suggested that the presence of a bar and the early cessation of star formation within a galaxy are intimately linked.
These results suggest possible role of bars in regulating and eventually quenching of star formation in galaxies. The suppression of star formation aided by bars within the region between the nuclear/central sub-kpc region and the ends of the bar is known as bar-quenching. There are two possible scenarios which can lead to bar-quenching. At the early stages of its formation, stellar bar induces torque on the gas within its extent which drives gas inflows to the centre of the galaxy. This can enhance star formation along the bar. But as the bar reaches its maximum length, the infalling gas is all consumed. This leads to depletion of gas and quenching of star formation within the central region of the galaxy (\citealt{combes1985spiral, spinoso2016bar, 2009A&A...501..207J,2015MNRAS.450.3503J, james2018star, 2019MNRAS.489.4992D,george2020more, 2020MNRAS.492.4697N, geron2021galaxy, 2024A&A...687A.255S}). Many observations and simulations suggest this scenario of bar-driven quenching of star formation, with bar hosting region to be devoid of H$\alpha$ and molecular gas (\citealt{gavazzi2015role, 2017A&A...598A.114C}). Another possible mechanism of bar quenching is due to the bar induced shocks and shear, which increase the turbulence of gas in the bar-region and in turn stabilise the gas against collapse leading to inhibition of star formation \citep{1998A&A...337..671R, haywood2016milky, khoperskov2018bar, 10.1093/mnras/stac3083, Kim_2024}. Several individual studies which focussed on the SFR, molecular gas fraction and star formation efficiency in the bar region, have found the star formation efficiency to be low, even with presence of molecular gas, and suggested that it could be due to shocks and shear produced by bar (\citealt{2022A&A...663A..61P, Maeda_2023}).  It is not well understood whether one or combination of these two processes, or a different unknown mechanism is responsible for bar-quenching. The net effect of both the processes explained above, is quenching of star formation within the co-rotation radius of barred galaxies, with little or no star formation in the sub-kpc nuclear region. The presence or absence of star formation in the central sub-kpc region may depend on the evolutionary stage of bar-quenching \citep{george2020more}. Other quenching mechanisms, such as cosmological starvation \citep{gavazzi2015role} and/or environmental effects \citep{2012MNRAS.423.1485S}, can transform these centrally quenched barred galaxies into fully passive galaxies \citep{2019A&A...628A..24G}.\\ 
The galaxies hosting strong bars are found to be gas poor and those which do not host a strong bar are found be be gas rich \citep{2012MNRAS.424.2180M}. This indicates that it might be difficult to form or grow bars in the discs of gas rich galaxies. Simulations do show that the presence of significant amount of gas can inhibit/slow down bar formation and bars form easily in gas poor discs (\citealt{Villa-Vargas_2010, athanassoula2013bar, 2017MNRAS.469.1054A}). Thus, an alternate explanation for the observed properties of barred galaxies, that they are globally redder, older and more metal rich than unbarred galaxies of similar mass, is that the strong bars are formed in quenched discs. However, strong bars do exist in gas rich discs galaxies \citep{2020MNRAS.492.4697N} and very recent studies have observed bar formation in gas-rich environments at redshifts of z $\sim$ 3.8 (\citealt{2024arXiv240401918A}). Thus, the role of bars in the suppression of star formation is still under inspection. 
Spatially resolved studies of barred galaxies to find the age of stellar populations across different regions of the galaxy will help to better understand the evolution of bar, coupling between the bar and disc evolution and hence the role of bars in the evolution of galaxies.  \\
The bar driven quenching scenario has been investigated in a spatially resolved manner by various studies (\citealt{gavazzi2015role, 2017A&A...598A.114C, james2018star, george2019insights, george2020more, 2024A&A...687A.255S} and references therein), using different tracers of star formation (such as molecular gas, atomic gas, dust emission, ionised gas, UV emission from young stars). All of the above studies are limited to nearby galaxies (10 -- 100 Mpc distance or redshift $<$ 0.02). 
From the study of spatially resolved star formation in local barred galaxies (z $\le$ 0.15), using MaNGA data, \cite{2020MNRAS.495.4158F} found that bars play an important role in the galaxy's ongoing star formation. However, due to MaNGA's limited field of view, they mainly analysed the bar region of the sample galaxies and does not include the extended disc region. Recent JWST observations (\citealt{Guo_2023, 10.1093/mnras/stae921}) have revealed barred galaxies at very high redshifts (z $\ge$ 2.5) suggesting that bar-driven quenching may also operate in the early Universe. And if this mechanism is effective at these redshifts, it could lead to galaxies being quenched very early when the Universe was still young. Hence, to better understand how passive galaxies form over time and the role played by bars in quenching star formation, we need to study barred galaxies in a spatially resolved manner (including the bar and disc regions) across different redshifts.
In this context, we analyse a sample of nearly face-on barred galaxies, which are centrally quenched and host star-forming discs, within a redshift range of 0.02 -- 0.06, using spatially resolved UV-optical colour-colour maps to understand the role of bars in quenching star formation.\\
The centrally quenched systems are identified based on their location in the SFR vs M$_{\star}$ plots created using the  M$_{\star}$ and SFR values from the Max Planck Institute for Astrophysics (MPA) and Johns Hopkins University (JHU) value-added catalogue (MPA - JHU VAC, \citealt{brinchmann2004physical}) and the GALEX-SDSS-WISE Legacy Catalogue (GSWLC, \citealt{2018ApJ...859...11S}). In the MPA-JHU catalogue, the fiber-based SFRs are derived from the SDSS spectra (the spectra is collected with 3 arcsec size fiber) using the H$\alpha$ line flux for star-forming galaxies, and an empirical relation between specific SFR and D$_{n}$4000 index for AGN or weak emission line galaxies. Aperture corrections are then estimated by deriving out-of-fiber SFRs via spectral energy distribution (SED) fitting to the SDSS u,g,r,i and z photometry. The GSWLC catalogue SFR estimates are based on the SED fitting from Ultraviolet (UV) to Mid Infrared (MIR) photometric data that fully covers the galaxy with no aperture correction required. \cite{cortese2020xgass} suggested that the aperture-corrected SFR estimates from the MPA - JHU SDSS DR7 catalogue, may not provide a fair representation of the global SFR of galaxies with extended star-forming discs. Thus, galaxies which are classified as quenched/passive in the M$_{\star}$ - SFR plane, based on the parameters from the MPA-JHU catalogue, but are non-passive (green valley or star-forming) based on the parameters from the GSWLC catalogue, are potential candidates for centrally quenched systems, with star-forming discs \citep{george2021bar}. These galaxies might be undergoing internal quenching, potentially driven by mechanisms like AGN feedback, influence of stellar bar or bulge as discussed before.
If the central quenching is due to stellar bar, then we expect to see older population in a region with size comparable to the size of the bar and younger population in the outer disc. We use the UV-optical colours of different regions of barred galaxies to probe the role of bar in quenching.\\ 
The UV emission in galaxies is predominantly from young star-forming regions, which host massive O and B type stars. However,  evolved, hot stellar populations such as the
horizontal branch stars, post asymptotic giant branch stars, and white dwarfs can emit in UV and this excess UV emission (in FUV) due to old and hot (or extreme) horizontal branch stars is known as UV upturn. We note that the AGN/LINER activity can also contribute to the UV emission in the galaxies, but in this work (as shown in Sect. \ref{sec:SS}) we study galaxies which do not host AGN. The UV-optical colour-colour map, introduced by \cite{2011ApJS..195...22Y}, which makes use of two colour indices, FUV$-$NUV and NUV$-$r, is a powerful tool to understand the nature of UV emission in galaxies and classify them into UV upturn, UV weak or star-forming categories. Along with the UV-optical colour-colour map, we also analyse the radial profiles of NUV$-$r colour, which is a good photometric indicator of star formation in the last 1 Gyr (Salim et al. 2005; \citealt{2007ApJS..173..619K,2014SerAJ.189....1S,Pan_2016}), of our sample galaxies. \\ 
The main goals of this study are: 
1) Identify barred galaxies which are probable candidates of centrally quenched systems, based on their location in the SFR vs  M$_{\star}$ plots created using the MPA-JHU and GSWLC catalogues, (2) Make spatially resolved UV-optical colour-colour maps of different regions of these galaxies and classify the regions as either star-forming or UV-upturn (due to old hot stellar population). 
Based on these colour-colour maps and the NUV$-$r colour profiles, check whether these systems are actually centrally quenched galaxies or not and to identify the location of star-forming region of the galaxies, (3) If the galaxies are indeed centrally quenched galaxies, then to check whether there is any correlation with the size of the centrally quenched region and the size of the bar, supporting the role of bar in quenching the inner regions of the galaxies. The structure of paper is as follows. Section \ref{sec:SS} describes the procedure followed for sample selection. Sections \ref{sec:data} and \ref{sec:ana}  provide information about data and the methods employed for the analysis of the whole sample. In Section \ref{sec:result}, we present our results regarding significant role of bar in quenching and try to understand the impact of different morphological features and their properties in central quenching. In Section \ref{sec:dis}, we discuss our results and try to understand the impact of different morphological features and their properties in central quenching and in Section \ref{sec:summary}, we summarise the work. We adopt standard flat cosmological model with $ \mathrm{H_{0}} = 70\; \mathrm{Km \;s^{-1}\; Mpc^{-1}}, \Omega_{M}= 0.3,  \; \text{and} \; \Omega_{\Lambda} = 0.7$. 

\section{Sample selection}
\label{sec:SS}
\begin{figure}
\centering
\begin{subfigure}{0.45\textwidth}
   \includegraphics[width=\linewidth]{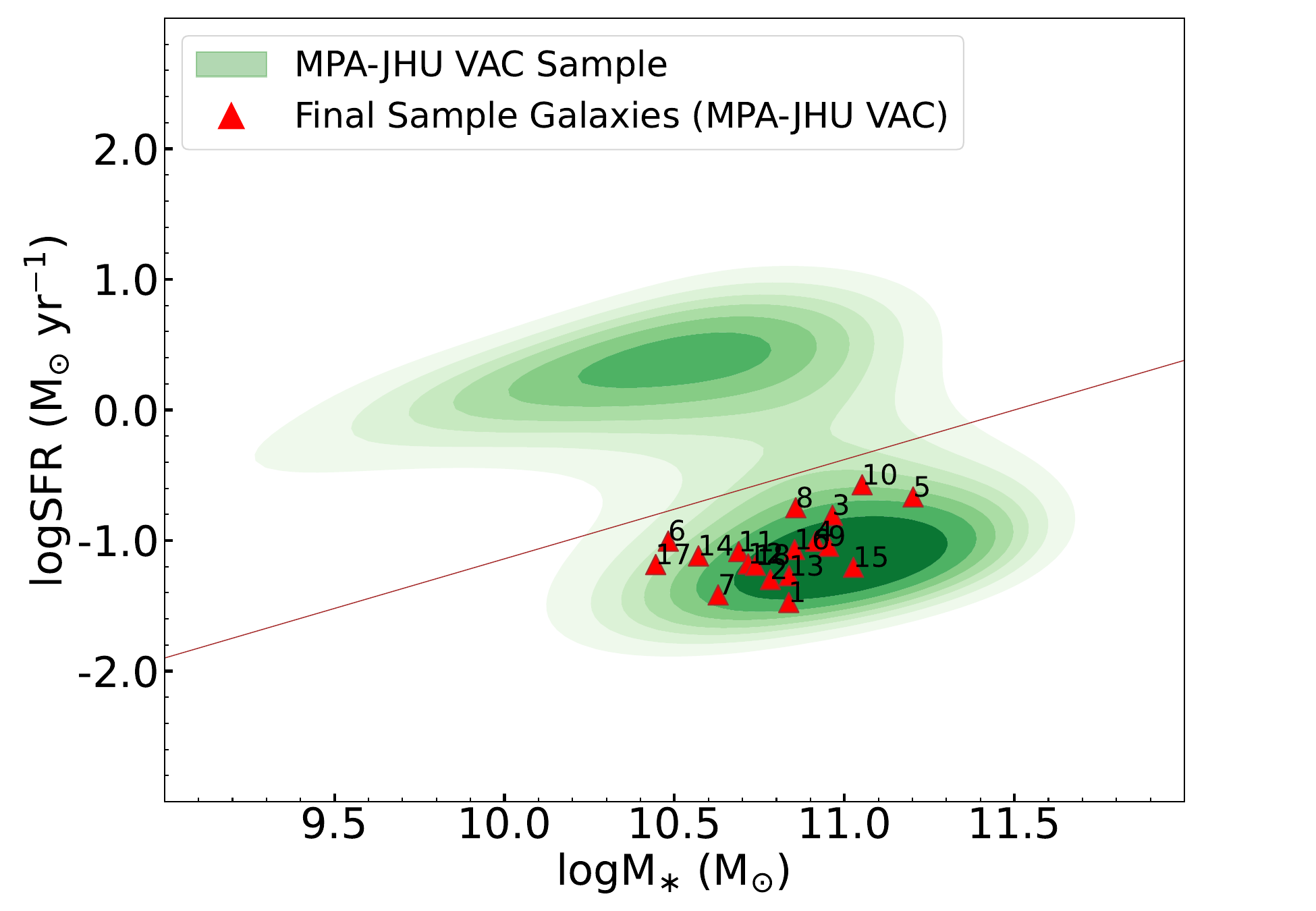}
   \caption{SFR-M$_{*}$ relation defined for MPA-JHU Value Added Catalogue \citep{brinchmann2004physical} (green contours) with red line representing SFMS$-$1.1 dex adopted from \cite{2020MNRAS.499..230B}. Our sample galaxies are quenched based on MPA-JHU SFR estimates.} \label{fig:x_a}
\end{subfigure}
\par\medskip
\begin{subfigure}{0.45\textwidth}
   \includegraphics[width=\linewidth]{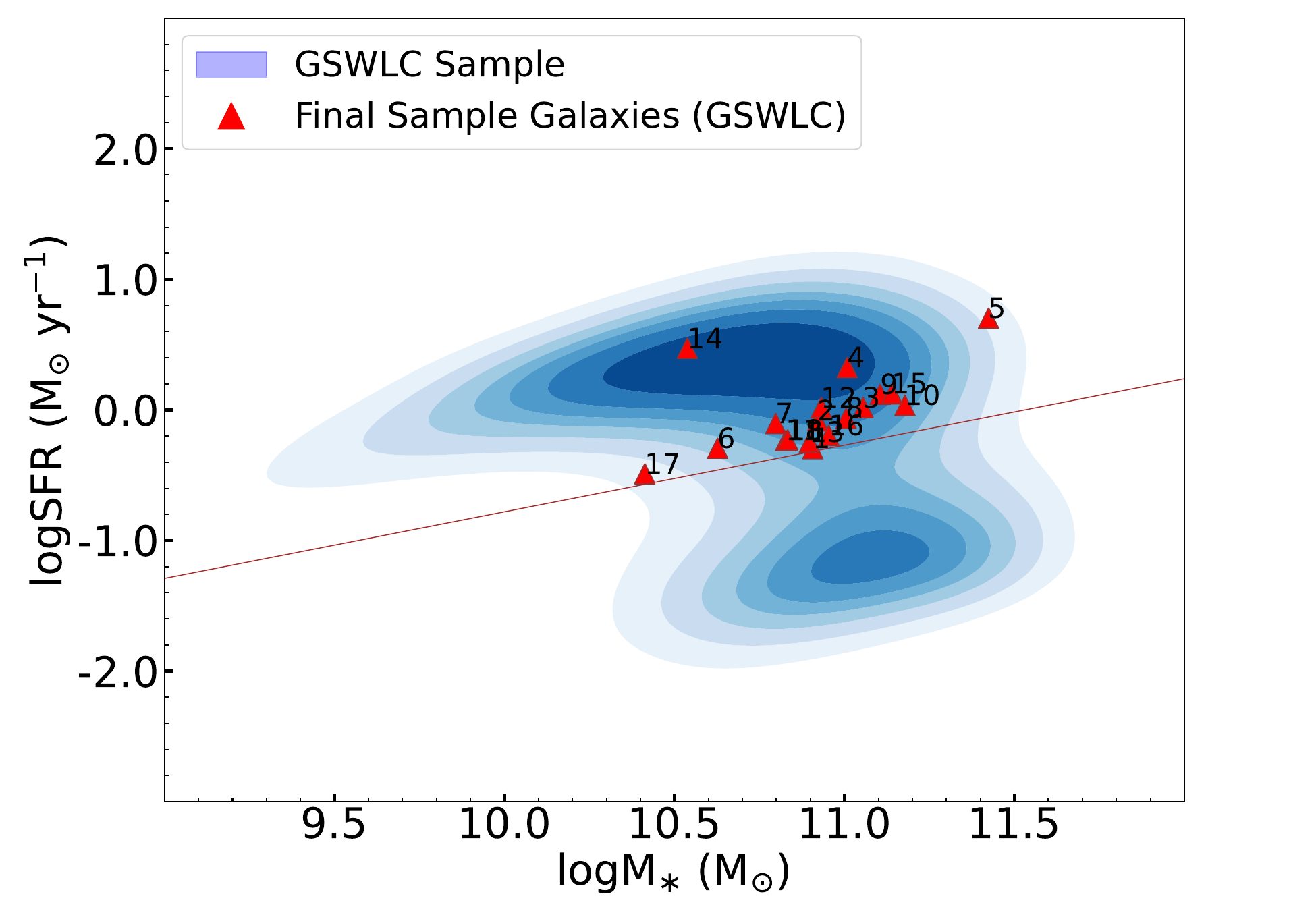}
   \caption{SFR-M$_{*}$ relation for GSWLC catalogue \citep{2018ApJ...859...11S} (blue contours) with red line representing SFMS$-$0.6 dex taken from \cite{Guo_2019}. Our sample galaxies are unquenched based on GSWLC SFR estimates. } \label{fig:x_b}
\end{subfigure}
\caption{ SFR-M$_{*}$ relation defined for the final sample of 18 barred galaxies with two different SFR indicators - MPA-JHU Value Added Catalogue and GSWLC M2 catalogue. In (a) the whole sample of 18 galaxies is lying in quenched region while in (b) they are lying above quenched region indicating these are non-passive systems. The discrepancy is due to the estimation of SFR using different techniques adopted for both catalogues as described in Sect. \ref{sec:SS}.}
\label{fig:1}
\end{figure}

 We used the catalogues provided by \citet{nair2010fraction} (hereafter, NA10) and \citet{kruk2018galaxy} (hereafter, K18) to identify the sample of barred spiral galaxies. A control sample of unbarred galaxies is also taken from NA10 and is discussed in a later Section (Sect. \ref{sec:dis}).\\
The catalogue by NA10 provides detailed visual morphological classifications for a comprehensive set of 14,034 galaxies from SDSS Data Release 4 (DR4). The catalogue gives information about a galaxy's morphology in terms of its T-type and whether it contains a bar. The T-type offers a scale (ranging from -5 to 0 for ellipticals and lenticulars, and from 1 to 7 for spirals, with higher values for Magellanic and irregular types) to quantify a galaxy's morphological type in a more continuous manner compared to the traditional Hubble classification. We selected spiral galaxies with T-types ranging from 1 to 7. From this catalogue, we identified 2,113 barred spiral galaxies. To select only those galaxies which are not significantly affected by external influence, we excluded those which are in pairs or undergoing interactions based on the given flags in the NA10 catalogue. This reduces the number of barred galaxy sample to 1567. Additionally, this catalogue also provides information on the type of AGN, based on the flux measurement from the SDSS spectra, and criteria from \citet{kauffmann2003stellar} and \citet{2001ApJ...556..121K}. 
We note that the feedback from AGN  can also quench star formation in the inner regions of the galaxies. In this study, we mainly focus on the role of the bar in quenching star formation. So we removed galaxies which are classified as AGN, and the sample reduced to 1044 barred galaxies. \\
The catalogue by K18 comprises of 2D photometric decomposition of $\sim$ 3461 local barred galaxies from the Galaxy Zoo 2 project (\citealt{2013MNRAS.435.2835W}). In their study, 2D decomposition on the SDSS images were performed with bar and disc components considered for all the galaxies, and a bugle component is used when required for a good fit. As we plan to study the spatially resolved colour-colour maps of our sample galaxies, we removed inclined galaxies with b/a $<$ 0.5 and also applied a cut-off in the redshift and removed those with z $>$ 0.06. As such after applying these two cuts, the number of barred galaxies reduced to 830 and 3327 barred galaxies from NA10 and K18, respectively resulting in 4157 total barred galaxies.\\ 
In order to identify the centrally quenched galaxies, we need to compare the nature of their star formation based on the parameters from both the MPA - JHU and the GSWLC catalogues. So, we cross-matched the sample of 4157 barred galaxies with both these catalogues. We found 2065 galaxies (491 from NA10 \& 1574 from K18) in common with defined logM$\ast$ and logSFR parameters. The MPA - JHU catalogue also provides information about the AGN class based on emission line diagnostics from the SDSS spectra. We removed galaxies which are classified as AGN or LINERs. This led to a sample of 1501 barred galaxies (376, NA10 \& 1125, K18) barred galaxies. The M$_{\star}$ - SFR  plots, created based on the parameters from the MPA - JHU and the GSWLC catalogues are shown in the upper and lower panels of Fig. \ref{fig:1}, with the classification lines taken from  \cite{2020MNRAS.499..230B} and \cite{Guo_2019} respectively.\\
\citet{2020MNRAS.499..230B} used the star-forming main sequence (SFMS) relation provided by \citet{renzini2015objective} and classified galaxies within the  $\pm$0.5 dex lines as star-forming galaxies, galaxies above the $+0.5$ dex line are considered as starburst galaxies, galaxies between $-0.5$ dex and $-1.1$ dex are considered as Green Valley galaxies, and galaxies below the $-1.1$ dex line (shown as red line in Fig. \ref{fig:1}a) are considered as passive galaxies. \cite{Guo_2019} found the MS relation using the sample from the GSWLC catalogue and classified galaxies within the $\pm$0.2 dex of the relation as star-forming, galaxies above the $+0.2$ dex line as starbursts, galaxies between -0.2 dex and $-0.6$ dex are considered as green valley galaxies, and galaxies below the $-0.6$ dex line (shown as red line in Fig. \ref{fig:1}b) are considered as passive galaxies. Thus the empirical form of the SFMS relation can be used to separate different regions on the M$_{\star}$ - SFR plane as star-forming, starburst, and passive. The cut of $-0.6$ dex below the MS to identify passive galaxies, is higher than the conventional cut used in literature to classify passive galaxies. So by selecting galaxies above this cut, we select the highly probable non-passive galaxies. We employed the above-described criteria to the sample of 1501 barred galaxies and selected those galaxies which are located below the red line (SFMS$-1.1$ dex corresponding to passive region in MPA - JHU catalogue) based on MPA - JHU catalogue SFR estimates (Fig. \ref{fig:1}a) but lie above the red line (SFMS$-0.6$ corresponding to passive region in GSWLC catalogue) based on GSWLC SFR estimates (Fig. \ref{fig:1}b). We found 87 galaxies (11, NA10 and 76, K18) galaxies out of 1501 that are classified as quenched according to the MPA - JHU parameters and classification provided by \citet{2020MNRAS.499..230B} but are classified as unquenched according to GSWLC parameters and the classification provided by \cite{Guo_2019}. These 87 barred galaxies are potential candidates of centrally quenched galaxies with central quenching likely due to the effect of bars. We then applied a stellar mass cut of M$_{\ast} > 10^{10} $M$_{\odot}$ to the sample to select massive galaxies and the sample reduced to 80 galaxies. We then checked for the availability of GALEX DR6/DR7 imaging data in both FUV and NUV bands for our sample of 80 barred galaxies ensuring full coverage. As we want to analyse the UV emission in the inner as well as outer regions of our sample galaxies, for further analysis, we selected only those galaxies which are observed under the Medium/Deep imaging surveys of the GALEX. These observations have a typical exposure time greater than 1000s.
This resulted in a subset of 39 galaxies (7, NA10 and 32, K18) with GALEX data. After identifying 5 duplicate galaxies present in both samples and removing them from the NA10 sample, our refined sample comprised of 34 galaxies, with 2 from NA10 and 32 from K18. \\
Since our goal is to study centrally quenched galaxies with no traces of ongoing (10 Myr) star formation, we visually inspected the SDSS spectra, which correspond to the central 3 arcsec region of these candidate galaxies, and removed those that show some amount of H$\alpha$ emission (which might indicate ongoing star formation or LINER activity). We found 12 galaxies that show emission in the central regions, which could contain those galaxies that are showing the last event of star formation in the sub-kpc regions and are in the final stages of bar-quenching. This sample will be analysed in detail in our future work. After removing the 12 galaxies which show emission in the central regions, our sample reduced to 22 galaxies, the most probable candidates of centrally quenched barred galaxies. We further cross-checked our final sample using the GSWLC M1 (\citealt{2016ApJS..227....2S}) based parameters for unquenched galaxies, retaining only those unquenched in both GSWLC M1 and GSWLC M2. The GSWLC M1 \citep{2016ApJS..227....2S} and GSWLC M2 \citep{2018ApJ...859...11S} differ in terms of their SFRs derived from SED fitting with the former used UV/optical photometry and the latter employed UV+Optical+mid-IR photometry. Additionally, the slope of the attenuation curve and the strength of the 2200 {\bf $\AA$} feature are taken as free parameters in GSWLC M2. Thus, our final sample consists of 18 barred galaxies, with 2 from NA10 and 16 from K18. The different criteria applied for the selection of the final sample are given in Table \ref{table:1}. The detailed information about the various properties of our sample galaxies is provided in Table \ref{table:2}. Majority of our sample galaxies have stellar masses higher than that of L$\ast$ galaxies. Three galaxies in our sample have HI mass estimates available from the ALFALFA-SDSS galaxy catalogue \citep{2020AJ....160..271D} and these HI masses are roughly 10\% of their stellar masses. 
The location of the final sample of 18 galaxies in the M$_{\star}$ - SFR  plane based on the MPA - JHU and the GSWLC catalogues are shown, as red triangles, in the upper and lower panels of Fig. \ref{fig:1}.\\
We note that we have not given any specific selection cuts to the K18 sample, to understand their environment. To evaluate the environment of these galaxies, we defined a cubic box with a width of 1 Mpc centred on each galaxy and estimated the number of neighboring galaxies within this volume using data from the NASA Extragalactic Database (NED\footnote{\url{https://ned.ipac.caltech.edu}}). Approximately 66$\%$ of the sample galaxies have no neighbours, 22$\%$ have one or two neighbours, and only 11$\%$ have 3 neighbours. The presence of nearby neighbours can influence the formation of bars in galaxies (\citealt{2009ApJ...702.1250R, Kazantzidis_2008}), but their contributions to the quenching of star formation in barred regions may be minimal. The impact of the environment is likely to primarily affect the disc star formation. Furthermore, studies indicate that environmental effects are nominal for galaxies with stellar masses exceeding 10$^{10.5}$ M$_{\odot}$ \citep{Guo_2019}.
As the aim of our study is to understand the effect of bars in the internal quenching of star formation, we do not expect the environmental effects to affect our final results.
\begin{table}[h!]
\caption{The final sample of barred galaxies is obtained based upon the following cuts and criteria mentioned in the table below from two catalogues, K18 and NA10. More details are described in Sect. \ref{sec:SS}.}
\label{table:1}
\centering
\resizebox{1.0\linewidth}{!}{
\begin{tabular}{|c|c|c|}
         \hline
         {\bf Sample Cuts} &   {\bf K18}  &  {\bf NA10 }\\ 
         \hline
         
        Spiral barred galaxies  &  3461  &  2113\\
         \hline
         Barred galaxies without pairs or interactions &  -  &  1567\\
         \hline
         Barred galaxies without AGN     & -  & 1044  \\
         \hline
         Barred galaxies (b/a $>$ 0.5) \& (z $<$ 0.06) &  3327  &   830 \\
         \hline
         Barred Galaxies in MPA-JHU and GSWLC  &  1574    &  491 \\
             (after removing undefined logM$\ast$ and logSFR) &   &  \\
         \hline
         Barred Galaxies without AGN and LINERs (BPT class $\leq$ 3) &  1125 &  376\\
         \hline
         Barred Galaxies quenched in MPA-JHU but unquenched in GSWLC & 76  & 11 \\
         \hline
         Barred galaxies with M$_{\ast}$ $>$ 10$^{10}$M$_{\odot}$ & 70 & 10 \\
         \hline
         Available GALEX MIS data (Exposure $>$ 1000s)  & 32 & 7 \\
         \hline
         Removing duplicate galaxies (present in both K18 \& NA10) from NA10  & 32 & 2 \\
         \hline
         Removing galaxies with H$\alpha$ emission in central 3" fibre sdss spectra &   20  &  2 \\
         \hline
         Barred Galaxies present in both GSWLC M1 and GSWLC M2 & 16 &  2 \\
         \hline
         Final Barred Galaxies Sample    &    \multicolumn{2}{c|}{18} \\
         \hline
         \end{tabular}
         }
\end{table}
\begin{figure*}[h!]
    \includegraphics[width=1.0\linewidth]{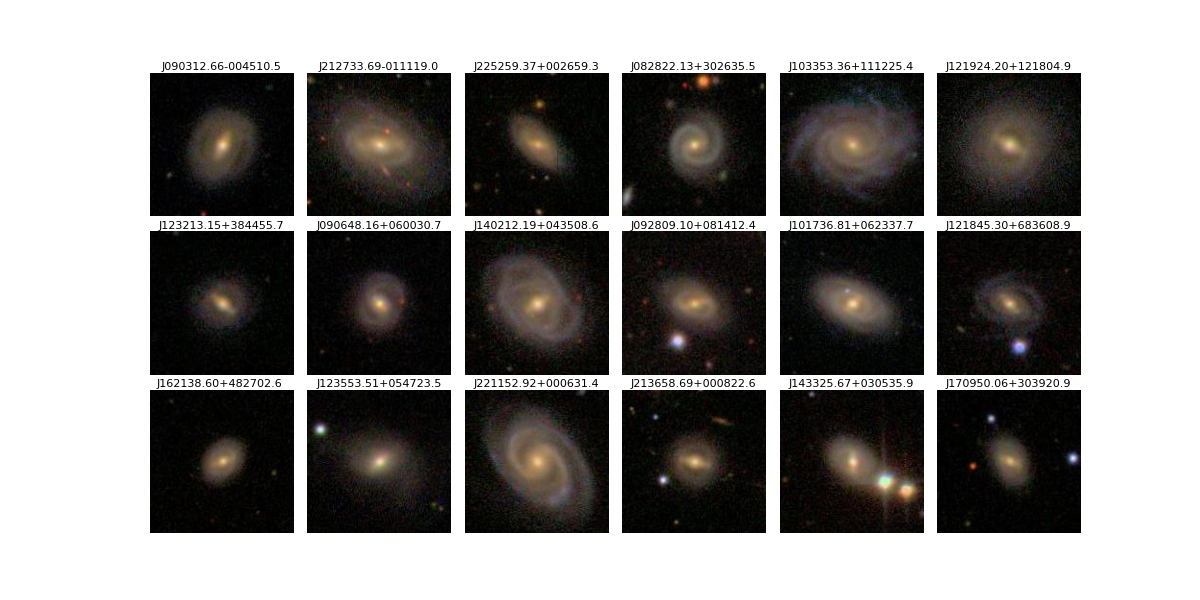}\
    \caption{Postage stamp of colour composite images of the final sample for 18 barred galaxies from SDSS DR7 u,g,r,i, and z bands. Each image is $\sim$ 1 arcmin $\times$ 1 arcmin with redshift ranging from 0.01- 0.06. The SDSS id is also provided on the top of each image. More information on the sources is available in Table \ref{table:2}.}
    \label{fig:2}
\end{figure*}
\begin{table*}[ht]
\caption{Details of each galaxy is mentioned. The angular to length scale is provided for different redshifts. The R$_{25}$ scale length in arcsec is taken from NED. Galaxy 4 and 5 are from NA10 and others are from K18. The major-to-minor axis (b/a) ratio is taken from respective catalogues. The stellar mass is given by GSWLC M2 catalogue. The HI mass is obtained from the ALFALFA-SDSS galaxy catalogue \citep{2020AJ....160..271D}. The bar length given here is estimated in our analysis.}
\label{table:2}
\centering
\begin{tabular}{llllllllll}
\hline \hline \\
\begin{tabular}[c]{@{}l@{}}Galaxy\\ Number\end{tabular} & \begin{tabular}[c]{@{}l@{}}RA\\ \\ (deg)\end{tabular} & \begin{tabular}[c]{@{}l@{}}DEC\\ \\ (deg)\end{tabular} & \begin{tabular}[c]{@{}l@{}}Redshift\\ \\ z\end{tabular} & \begin{tabular}[c]{@{}l@{}}Scale\\ \\ (kpc/arcsec)\end{tabular} & \begin{tabular}[c]{@{}l@{}}R$_{25}$\\ \\ (arcsec)\end{tabular} &  b/a & \begin{tabular}[c]{@{}l@{}}Stellar Mass\\ \\ $10^{10}M_{*}$\end{tabular} & \begin{tabular}[c]{@{}l@{}}HI Mass\\ \\ $10^{10}M_{*}$\end{tabular} &\begin{tabular}[c]{@{}l@{}}Bar Length\\ \\ (kpc)\end{tabular} \\ \\ \hline \\

1 &135.80276 & -0.752927 & 0.0483   & 0.946 & 20.86 & 0.76 & 10.91 & - &  15.19     \\\vspace{0.1cm}
2 &321.89042 & -1.188618 & 0.0304   & 0.608 & 34.03 & 0.68 & 10.92 & -&  13.75    \\\vspace{0.1cm}
3 &343.24741 & 0.449791  & 0.0521   & 1.016 & 20.10  & 0.52 & 11.06 & -&  17.16    \\\vspace{0.1cm}
4 &127.09223 & 30.443203 & 0.0545   & 1.060  & 17.79 & 0.95  & 11.01 & -&   10.16   \\\vspace{0.1cm}
5 &158.47236 & 11.207063 & 0.0498   & 0.974 & 32.65 & 0.63 & 11.42 & 10.62 &  13.93   \\\vspace{0.1cm}
6 &184.85086 & 12.301379 & 0.0263   & 0.529 & 26.79 & 0.95 & 10.63 &  09.68 & 9.03   \\\vspace{0.1cm}
7 &188.05482 & 38.748826 & 0.0526   & 1.025 & 15.18 & 0.92  & 10.80 & -&  12.17   \\\vspace{0.1cm}
8 &136.70071 & 6.008539  & 0.0592   & 1.145 & 15.34 & 0.86  & 11.00 & -&  17.23    \\\vspace{0.1cm}
9 &210.55081 & 4.585725  & 0.0403   & 0.797 & 28.99 & 0.79  & 11.10 & 10.09 & 13.26   \\\vspace{0.1cm}
10 &142.03791 & 8.236787  & 0.0573   & 1.111 & 19.42 & 0.69  & 11.18 & -&  13.20    \\\vspace{0.1cm}
11 &154.40339 & 6.393795  & 0.0326   & 0.651 & 22.71 & 0.60   & 10.83 & -&  5.16    \\\vspace{0.1cm}
12 &184.68877 & 68.602475 & 0.0593   & 1.147 & 17.94 & 0.74  & 10.93 & -&  10.90    \\\vspace{0.1cm}
13 &245.41087 & 48.450751 & 0.0585   & 1.132 & 12.66 & 0.73  & 10.90 & -&   11.66   \\\vspace{0.1cm}
14 &188.97297 & 5.789857  & 0.0419   & 0.827 & 23.29 & 0.54  & 10.54 & -& 7.20     \\\vspace{0.1cm}
15 &332.97048 & 0.108692  & 0.0333   & 0.664 & 31.00    & 0.72  & 11.14  & -& 9.99     \\\vspace{0.1cm}
16 &324.24455 & 0.139626  & 0.0569   & 1.103 & 13.76 & 0.91  & 10.95 & -&  14.85    \\\vspace{0.1cm}
17 &218.35703 & 3.093298  & 0.0296   & 0.593 & 15.38 & 0.63  & 10.41 & -&  5.17    \\\vspace{0.1cm}
18 &257.4586  & 30.655814 & 0.0547   & 1.063 & 14.37 & 0.69  & 10.83 & - &  8.42    \\
\hline
\end{tabular}%

\end{table*} 

\section{Data}
\label{sec:data}
Our aim is to study the spatially resolved UV-optical colour-colour maps of the centrally quenched barred galaxies. For this, we have used the SDSS r-band (5500-7000 \AA\ with central wavelength 6166 \AA) optical images from the SDSS DR7, \citealt{2009ApJS..182..543A}) for the photometric analysis. We opted for SDSS data over deeper optical surveys like DECaLS \citep{2019AJ....157..168D} because the two catalogues from which our sample is drawn are based on the SDSS database and provide structural properties using SDSS images exclusively. In SDSS r-band image, each pixel traces 0.396 arcseconds and the point spread function (PSF) in r-band is 1.32 arcsecond. The data is readily downloaded from SDSS DR7 skyserver. The postage stamp of SDSS colour composite images taken from SDSS skyserver for all our sample galaxies is shown in Fig. \ref{fig:2}. We have also employed GALEX (Galaxy Evolution Explorer, \citealt{2005ApJ...619L...1M}) FUV (1344–1786 \AA) and NUV (1771–2831 \AA) bands imaging data with effective wavelengths, 1538.6 and 2315.7 \AA\ respectively and they are accessed through the MAST\footnote{\url{https://mast.stsci.edu/portal/Mashup/Clients/Mast/Portal.html}} portal. The FWHM for FUV and NUV channels corresponds to 4.2 and 5.3 arcseconds respectively (\citealt{2007ApJS..173..682M}).\\
The image cutouts of size $\sim 2' \times 2' $ are made for our sample galaxies, centred on the source coordinates. The r-band images are background subtracted based on mean background calculated from different source-free positions in the images while the background maps are available for GALEX images for background subtraction. As the spatial resolution of the GALEX FUV ($\sim$ 4.2"), GALEX NUV ($\sim$ 5.3"), and SDSS r-band ($\sim$ 1.3") images are different, we first degraded the GALEX FUV and SDSS r-band images to the resolution of the GALEX NUV having coarser resolution to match the NUV PSF. 
The FUV and r-band images are degraded to the NUV resolution using Gaussian 2D convolution, where the width (defined as $\sigma$) of the Gaussian kernel used to convolve with the original images is calculated using target PSF and original PSF. The convolution is performed in pixel space for both FUV and r-band images, with $\sigma =$ 5.52 pixels (2.19 in arcsec) for the r-band image and $\sigma$ = 0.92 pixels (1.38 arcsec) for the FUV image. Foreground as well as nearby sources are all masked with the minimum pixel count. These background subtracted and cleaned image cutouts are used for further analysis. However, for the measurement of bar lengths, described in Sect. \ref{subsec:barlength}, we use the original SDSS r-band images with a native resolution of $\sim$ 1.3".

\section{Analysis}
\label{sec:ana}
\subsection{UV-Optical Colour-Colour Map}
\label{subsec:UV}
\begin{figure}[h!]   
   \centering
    \includegraphics[width=1.0\linewidth]{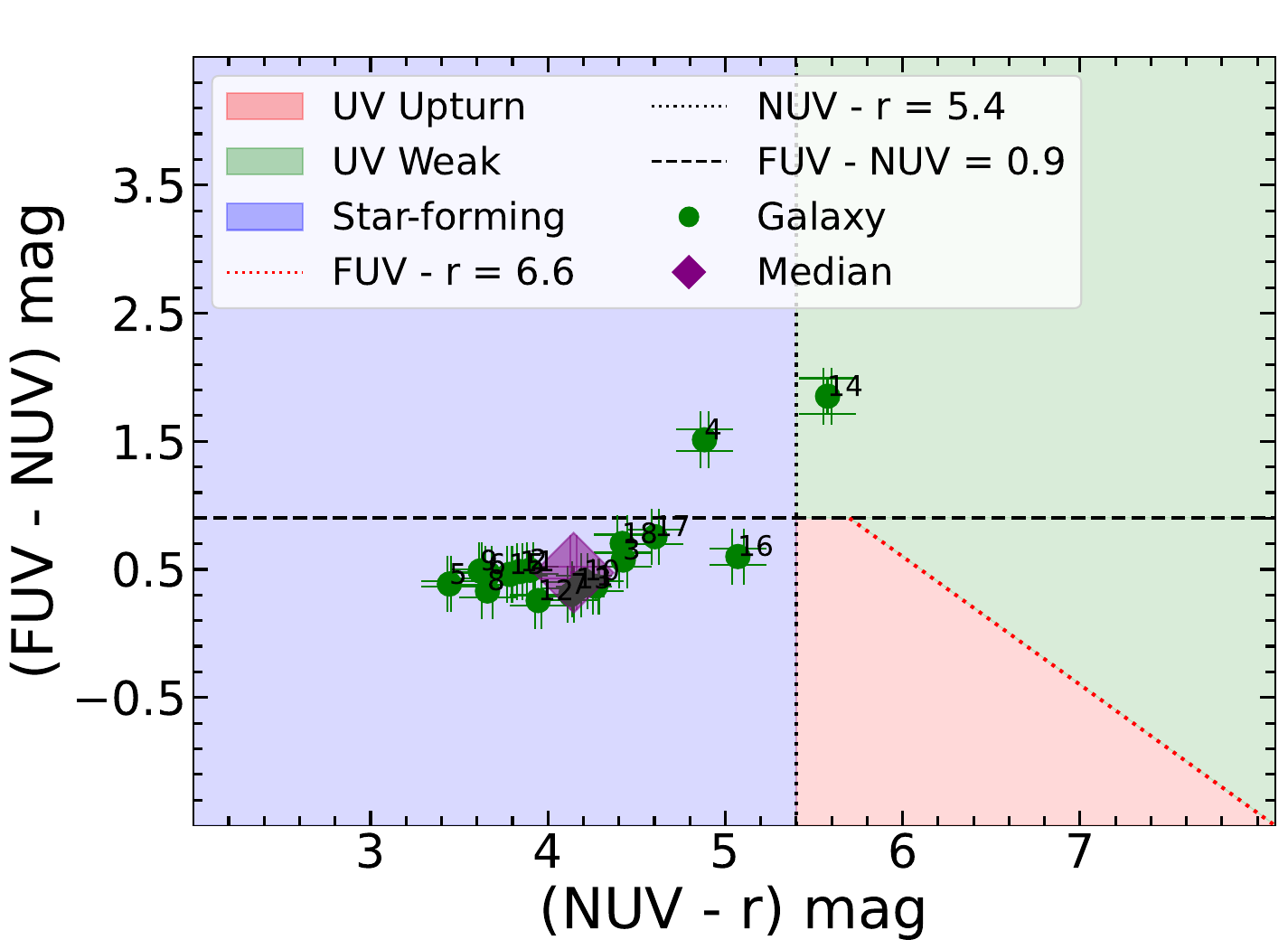}
\caption{The FUV$-$NUV and NUV$-$r colour colour map plotted with integrated colours (green points) of final sample (18 galaxies) upto R$_{25}$ scale length (radii at which surface brightness reaches 25 mag arcsec$^{-2}$ in the B-band). The median value for the sample is shown in purple. This UV-Optical colour-colour map is used to distinguish UV emission coming from both young stellar population and horizontal branch old stellar population. Out of 18 galaxies, 1 galaxy lie in UV weak and close to UV upturn while other 17 galaxies are lying in star-forming region. This signifies that the galaxies of our sample are indeed star-forming.}
\label{fig:3}
\end{figure}

Our sample of 18 barred spiral galaxies are found to be passive in nature based on their parameters in the MPA-JHU catalogue and are found to be non-passive based on their parameters in the GSWLC catalogue. As explained earlier, we leverage this discrepancy to identify them as the centrally quenched systems with outer star-forming discs. However, \cite{2021ApJ...911...57Z} noted that the SFR based on the GSWLC, could have been overestimated as the UV and IR emission of quenched galaxies can be contaminated by old hot stellar population and/or by AGN/LINER activity. We note that we have already removed AGN/LINER galaxies from our sample. To understand the contribution of UV emission from old hot stellar population, we used the UV-Optical colour-colour map criteria provided by \citet{2011ApJS..195...22Y}. The excess of UV emission (in FUV) due to old and hot horizontal branch stars is known as UV upturn. \citet{2011ApJS..195...22Y} investigated the UV upturn phenomenon in elliptical galaxies and used their foreground reddening corrected NUV$-$r, FUV$-$NUV and FUV$-$r colours to classify them as UV upturn galaxies, UV weak galaxies or those with residual star formation. The limit of NUV$-$r = 5.4 was used to differentiate UV emission arising from older hot stellar populations, such as horizontal branch (HB) and extreme horizontal branch (EHB) stars, from that produced by young stellar populations. Young massive stars emit intense UV radiation, resulting in a bluer NUV$-$r colour \citep{2011ApJS..195...22Y}. A FUV$-$NUV colour $<$ 0.9 is indicative of a rising UV slope with decreasing wavelength and corresponding to a flat UV spectrum in the $\lambda-F_{\lambda}$ domain, it is used to identify the UV upturn nature of galaxies, which reflects the presence of old hot stars. The criterion FUV$-$r $<$ 6.6, represents a high relative UV flux to optical flux. \\ 
The FUV, NUV, and r-band magnitudes of our sample galaxies are estimated using the fixed elliptical aperture photometry on the PSF matched and background subtracted FUV, NUV, and r-band images. We used the R$_{25}$ value (radii at which surface brightness reaches 25 mag arcsec$^{-2}$ in the B-band) given in Table \ref{table:2} (taken from NED) as the galaxy semi-major axis. We used the position angle extracted from the Ellipse task and b/a values given in the NA10 and K18 to define the elliptical apertures in the r, FUV, and NUV images. The count rates are converted to magnitudes using the appropriate zero points (18.82 for GALEX FUV and 20.08 for NUV \citep{2007ApJS..173..682M}; SDSS r-band photometric zero point is provided in image header along with atmospheric extinction parameters). The magnitudes are corrected for the foreground galactic dust extinction using the reddening values provided by \cite{2011ApJ...737..103S} and the reddening law by  \cite{1989ApJ...345..245C}. The reddening corrected colours of our sample galaxies are plotted on the colour-colour map, which is shown in Fig. \ref{fig:3}. Extinction errors along with magnitude errors are also taken into account in the colours of each galaxy.\\
From our initial sample of 18 galaxies, we identified one galaxy, galaxy no. 14, to be in the UV weak region of the UV-optical colour-colour map. The remaining 17 galaxies have UV-optical colours corresponding to ongoing star formation. However, as discussed earlier, the SDSS central spectra of these galaxies and their location in the M$_{\star}$ -- SFR  plot based on the parameters from the MPA-JHU catalogue, indicate that they are passive galaxies. These suggest that the ongoing star formation in these galaxies might be happening in their outer disc and these galaxies are centrally quenched. 
Indeed, also from the visual inspection of the  SDSS colour images shown in Fig. \ref{fig:2}, it is evident that these galaxies are characterised by a passive inner component (presence of regions with red colour) and a star-forming outer disc (relatively bluer colour in outskirts). As these galaxies do not host AGN, the quenching of star formation in the central regions might be due to bars and/or bulges. In the next sub-section, we discuss the bulge properties of our sample galaxies.
\subsection{Bulge properties}
\label{subsec:bulge}
Our sample contains barred galaxies contained from the catalogue of K18 and NA10. The K18 sample provides structural parameters for the barred disc galaxies using bulge+bar+disc decomposition in five SDSS bands (u,g,r,i, and z) by employing automated GALFITM tool which is a  modified version of GALFIT3.0 \citep{2010AJ....139.2097P}. Basically, GALFIT performs a two dimensional fitting  of a parametric function to the surface brightness profile for the galaxy. The image of the galaxy is fitted with an exponential or s\'ersic profile, and the model and residual image are obtained after convolving the model with the PSF of the image. It provides structural properties like photometric magnitude, effective radius (r$_{e}$), s\'ersic index (n), axis ratio (b/a), and position angle (PA) for each structural component. Out of 18 galaxies, 16 galaxies have their structural parameters available from K18. Two of our sample galaxies (Galaxy no. 4 and 5) are from NA10 and do not have the structural properties available. For these 2 galaxies, we used GALFIT tool on SDSS r-band image and obtained their structural parameters. For comparison, we performed a similar exercise on other galaxies with catalogue values and confirmed that the values we derived matched those in the catalogue. In Table \ref{table:3}, we provide the photometric r-band magnitude of bar (r$_{bar}$) and bulge (r$_{bulge}$), their effective radii (r$_{e,bar}$ and r$_{e,bulge}$), and s\'ersic indices (n$_{bar}$ and n$_{bulge}$). 
The s\'ersic index of our sample galaxies, excluding galaxy no. 14, is less than 2, indicating the pseudo nature of these bulges (\citealt{2008AJ....136..773F,2010ApJ...716..942F}). The galaxy no. 14, was found to be falling in the UV weak region in Fig. \ref{fig:3} and hence was already removed from our further analysis. The suppression of star formation is expected from classical bulges, which are formed during mergers. However, our sample does not have any galaxies with classical bulges. This suggests that the suppression of star formation in the central regions of our sample galaxies can be considered to be due to the action of bar.\\
If the central quenching is due to the effect of bar, then we expect to see no recent star formation in a region with size comparable to the size of the bar (\citealt{james2009halpha,george2019insights,george2020more}, and references therein). \cite{2024A&A...687A.255S} did a spatial level deviation of distance from the MS and showed that SFR is indeed very low along the bar region. To check these hypotheses, we want to identify the nature of star formation in different regions of these galaxies. We plan to do this by creating spatially resolved UV-optical colour-colour maps of these galaxies. In order to divide the galaxy into different sub-regions, especially with respect to the size of the bar, we estimated the length of the bar in these galaxies using isophotal analysis. This is described in the next Section.
\subsection{Measurement of bar length}
\label{subsec:barlength}
\begin{figure}
\centering
\begin{subfigure}{0.45\textwidth}
   \includegraphics[width=\linewidth]{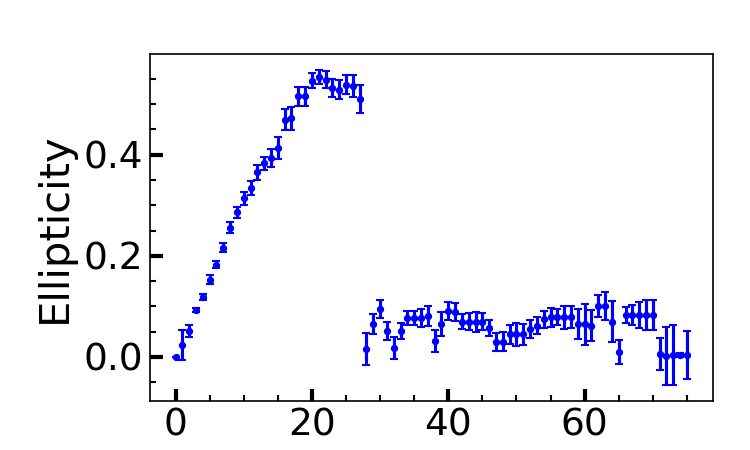}
   \caption{Ellipticity vs SMA } \label{fig:x_a}
\end{subfigure}
\hspace{0\linewidth} 
\begin{subfigure}{0.45\textwidth}
   \includegraphics[width=\linewidth]{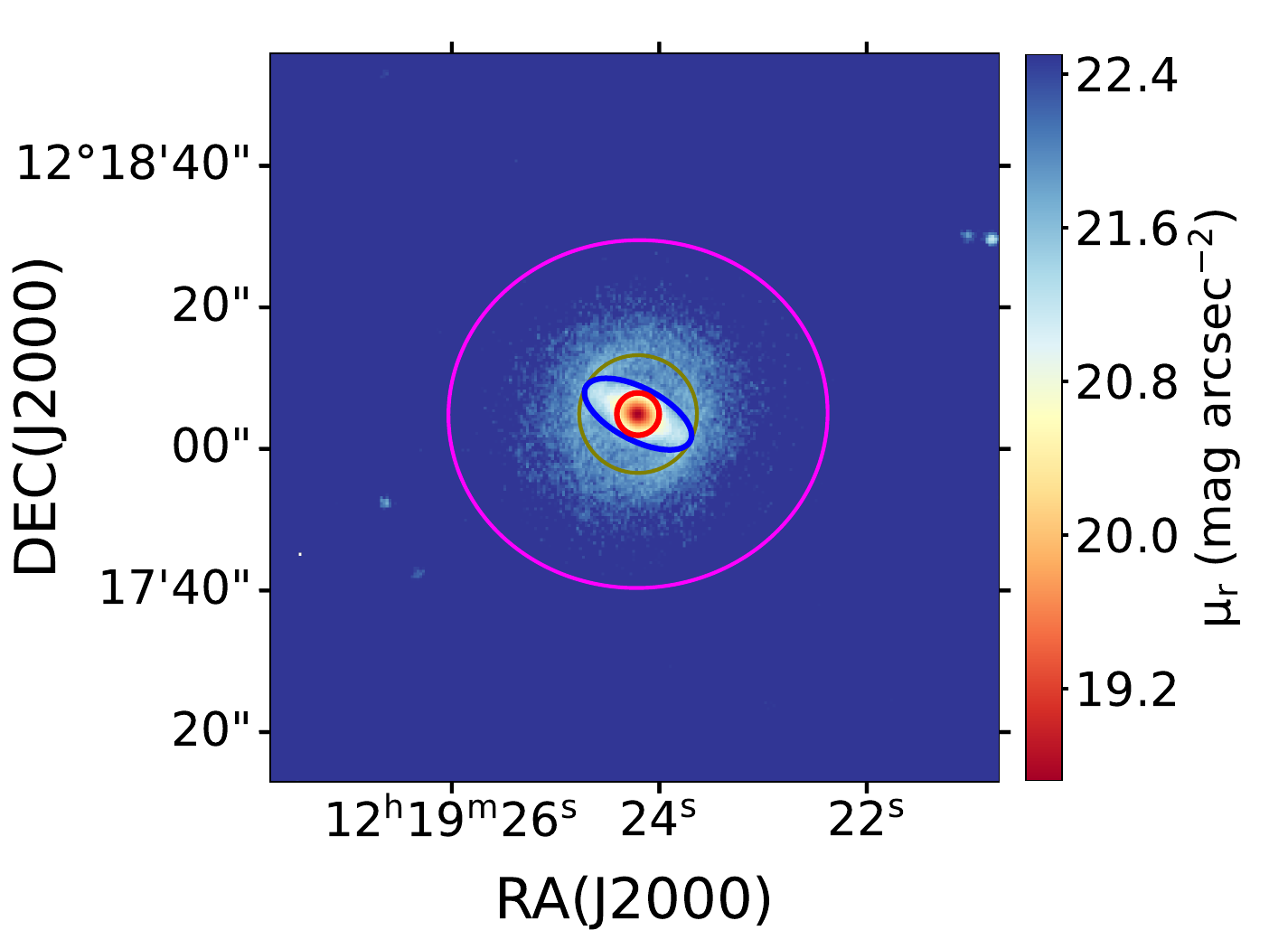}
   \caption{SDSS r-band} \label{fig:x_b}
\end{subfigure}
\caption{Fig. 4a shows the radial profile of ellipticity, where it rises as the radius increases, peaking at the bar length and then decreases matching the disc's ellipticity. Fig. 4b illustrates SDSS r-band image in terms of surface brightness $\mu_r$ (mag arcsec$^{-2}$) along with the various apertures used to represent distinct regions like bulge region (red), elliptical bar region (blue), circular bar region (brown), and annular disc region (magenta) for a galaxy.}
\label{fig:4}
\end{figure}

The bar length can be measured using different known methods like ellipse fits (\citealt{Jogee_2004, 10.1111/j.1365-2966.2005.09560.x, 2007ApJ...657..790M, Yu_2020,Guo_2023}), Fourier decomposition (\citealt{1990MNRAS.245..130A, 2020MNRAS.491.2547R,2022MNRAS.512.5339R}) or simply by visual inspection \citep{Cheung_2013}. Although there is no general agreement on which method is more preferable, each method comes with its advantages and limitations (\citealt{2009A&A...495..491A, Lee_2019}). The Fourier decomposition is very suitable for stronger bars due to high sensitivity of Fourier mode m=2 \citep{2002MNRAS.330...35A} but less effective for weak bars \citep{Lee_2019}. Similarly, visual inspection generates better outcomes only for stronger bars. Compared to these two methods, isophotal ellipse fits and the radial profile of ellipticity is more roboust to find bar length and applies for both strong and weak bars.  Isophotes, representing contours of constant surface brightness in a galaxy image, can reveal features such as bars, spiral arms, and other substructures. By fitting ellipses to isophotes at various radii, we can map a galaxy's two-dimensional light distribution in detail. This method only fails for low signal-to-noise and clumpy structures. In this paper, our sample of barred galaxies is drawn from K18 and NA10. The K18 galaxies are further a subsample of Galaxy Zoo 2 (GZ2) project (\citealt{2008MNRAS.389.1179L, 2013MNRAS.435.2835W}) where bars are identified based on a debiased bar likelihood and the K18 primarily selected strong and intermediate bars in their cuts. Similarly, NA10 employed visual identification for bars and mentioned that their bars identified belong to subclasses of the strong bar classification in RC3 \citep{1991rc3..book.....D}. Consequently, our sample galaxies have all been visually identified with structural properties avoiding the issues of low signal-to-noise and any clumpy structures. \\
To determine the bar length, we performed ellipse fitting on the original SDSS r-band images, (with a native resolution of 1.3") of our sample using the {\it Ellipse} task from the {\it photutils isophote} module\footnote{\url{https://photutils.readthedocs.io/en/stable/isophote.html}}. The ellipticity ($\epsilon$) profile typically peaks at the bar's extent and decreases as it transitions to the disc (see Fig. \ref{fig:4}a), while the position angle (PA) of the isophotes aligns with the bar's PA, resulting in a flat PA profile across the bar region.
We first located the centre by averaging the isophote distribution with $\epsilon$ and PA taken as free parameters and then fixing the centre to extract the $\epsilon$ and PA profiles. This method, as described by \citet{1987MNRAS.226..747J}, is widely established and frequently used for identifying bars (\citealt{1995A&AS..111..115W, 2007ApJ...657..790M, consolandi2016automated}) and determining their lengths. Also all the bars in our sample meet the criteria mentioned in \cite{Jogee_2004} with peak ellipticity at $\epsilon$  $\geq$ 0.3 and small PA variations ($\delta$ PA$_{bar} \lesssim$15$^{\circ}$). 
According to \citep{2004A&A...415..941E}, for reliable estimation of bar properties, the semi-major axis length of the bar should be larger than $\sim$ 2 $\times$ PSF FWHM. All the galaxies in our sample satisfy this criterion. 
We applied the above steps to study the bar regions across our sample galaxies. However, the caveats we found for galaxies at higher inclinations or relatively higher redshift, are that the $\epsilon$ profiles either smoothen out to the disc ellipticity or show multiple peaks (Galaxy No. 3, 5, and 11). In such cases, visual inspection was used to determine the bar lengths.

\begin{table}[h!] 
\captionsetup{width=1.0\linewidth}  
\caption{The table contains structural parameters from K18, derived from the photometric decomposition of barred disc galaxies in five SDSS bands (u,g,r,i, and z) using the automated GALFITM tool. For the two galaxies (marked with $\ast$) from NA10, we additionally used the GALFIT tool to obtain their structural parameters.}
\label{table:3}
\centering
\begin{tabular}{ccccccc}
\hline\hline
\multicolumn{1}{c}{Galaxy} & \multicolumn{1}{c}{\begin{tabular}[c]{@{}c@{}}r$_{bulge}$\\ \\ {[}mag{]}\end{tabular}} & \begin{tabular}[c]{@{}l@{}}r$_{e,bulge}$\\ \\ {[}kpc{]}\end{tabular} & n$_{bulge}$  & \multicolumn{1}{c}{\begin{tabular}[c]{@{}c@{}}r$_{bar}$\\ \\ {[}mag{]}\end{tabular}} & \multicolumn{1}{c}{\begin{tabular}[c]{@{}c@{}}r$_{e,bar}$\\ \\ {[}kpc{]}\end{tabular}} & n$_{bar}$    \\ \hline \\
1  & 16.84 & 0.85 & 0.87 & 17.75 & 3.43  & 0.25 \\ \vspace{0.1cm}
2 & 16.56 & 0.77 & 0.96 & 16.68 & 4.85 & 0.31 \\ \vspace{0.1cm}
3  & 19.21 & 0.75 & 0.27 & 18.33 & 3.26   & 0.76 \\ \vspace{0.1cm}
4$^*$   & 17.43   &  0.52  & 0.84  & 18.52  & 3.63 &  0.23\\ \vspace{0.1cm}
5$^*$ & 17.78  & 0.70 & 1.14 & 18.36 & 2.74  & 0.21 \\ \vspace{0.1cm}
6  & 16.80  & 0.57 & 0.86 & 17.42 & 2.92 & 0.40  \\ \vspace{0.1cm}
7  & 17.37 & 1.06 & 0.81 & 17.93 & 4.29 & 0.27 \\ \vspace{0.1cm}
8  & 17.57 & 0.65 & 0.63 & 17.01 & 2.64  & 0.52 \\ \vspace{0.1cm} 
9  & 16.67 & 0.62 & 0.93 & 16.90  & 3.10  & 1.10  \\ \vspace{0.1cm}
10  & 18.11 & 0.70 & 0.35 & 17.68 & 5.79 & 0.10  \\ \vspace{0.1cm}
11  & 16.75 & 0.39 & 0.82 & 17.29 & 2.17  & 0.48 \\ \vspace{0.1cm}
12 & 17.65 & 0.97 & 0.81 & 18.16 & 4.07  & 0.20  \\ \vspace{0.1cm}
13  & 18.24 & 0.75 & 0.59 & 18.74 & 5.41 & 0.23 \\ \vspace{0.1cm}
14  & 15.69 & 1.85 & 2.82 & 18.69 & 6.65 & 0.17 \\ \vspace{0.1cm}
15  & 16.69 & 0.72 & 0.8  & 16.77 & 4.06 & 0.21 \\ \vspace{0.1cm}
16  & 17.25 & 1.09  & 0.85 & 17.39 & 4.80    & 0.48 \\ \vspace{0.1cm}
17  & 16.91 & 0.38 & 0.36 & 17.06 & 1.94  & 0.45 \\ \vspace{0.1cm}
18  & 18.51 & 0.64 & 0.26 & 18.20  & 3.55  & 0.30 \\ \hline
\end{tabular}%
\end{table}

\begin{figure*}
\centering
\begin{subfigure}{0.28\textwidth}
   \includegraphics[width=\linewidth]{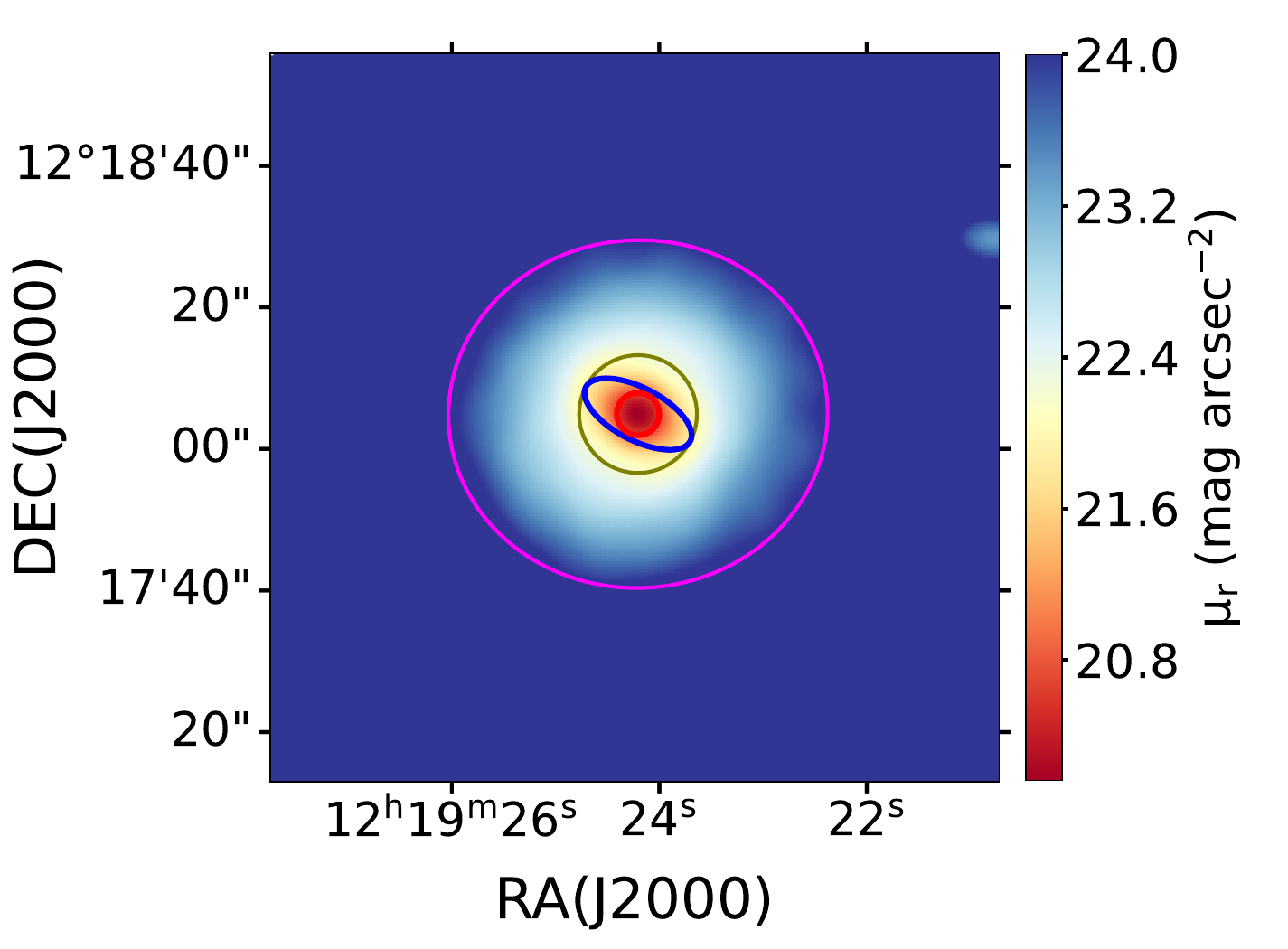}
   \caption{SDSS Degraded r-band} \label{fig:x_a}
\end{subfigure}
\hspace{0\linewidth} 
\begin{subfigure}{0.28\textwidth}
   \includegraphics[width=\linewidth]{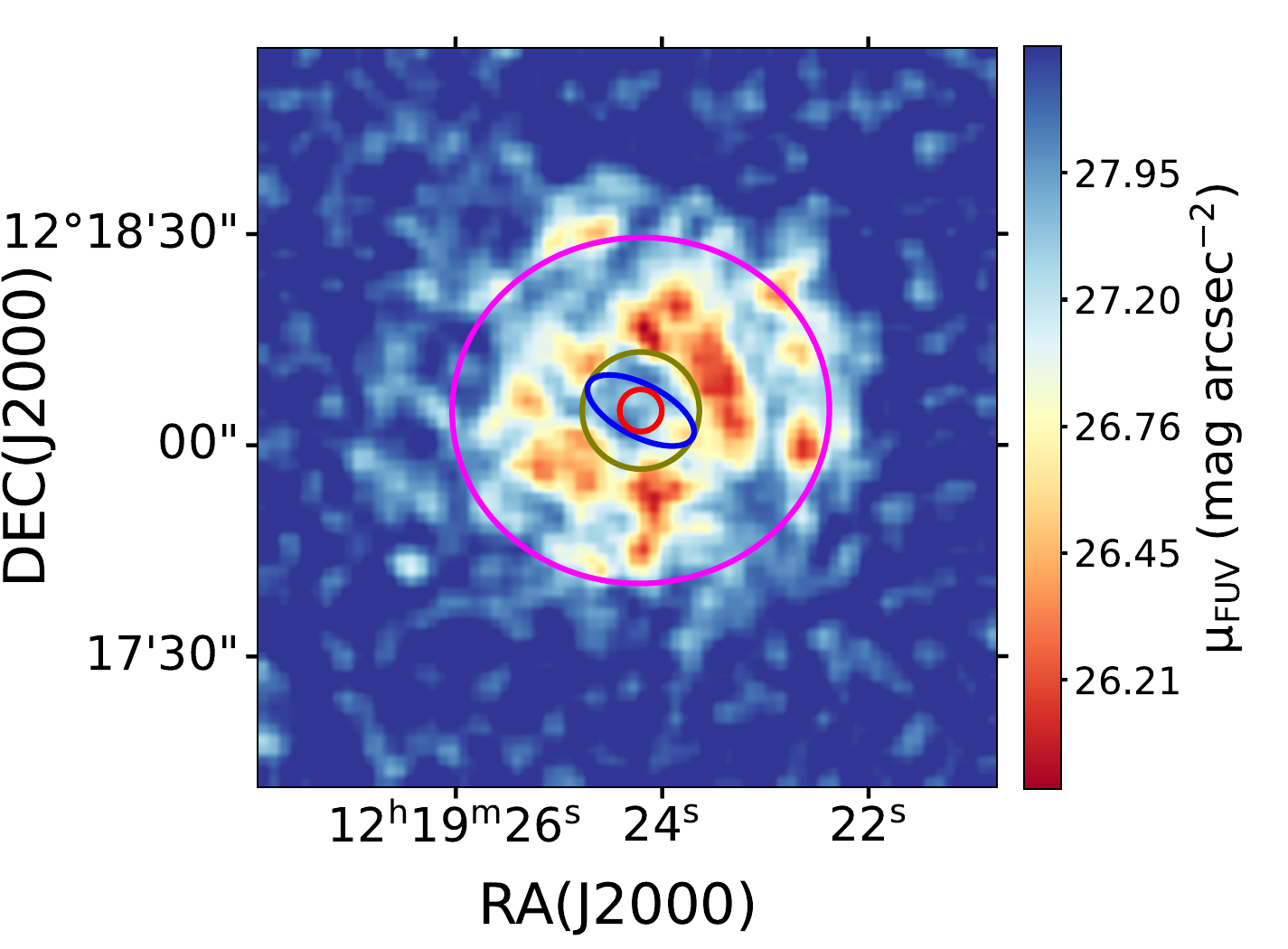}
   \caption{Degraded GALEX FUV} \label{fig:x_b}
\end{subfigure}
\hspace{0\linewidth} 
\begin{subfigure}{0.28\textwidth}
   \includegraphics[width=\linewidth]{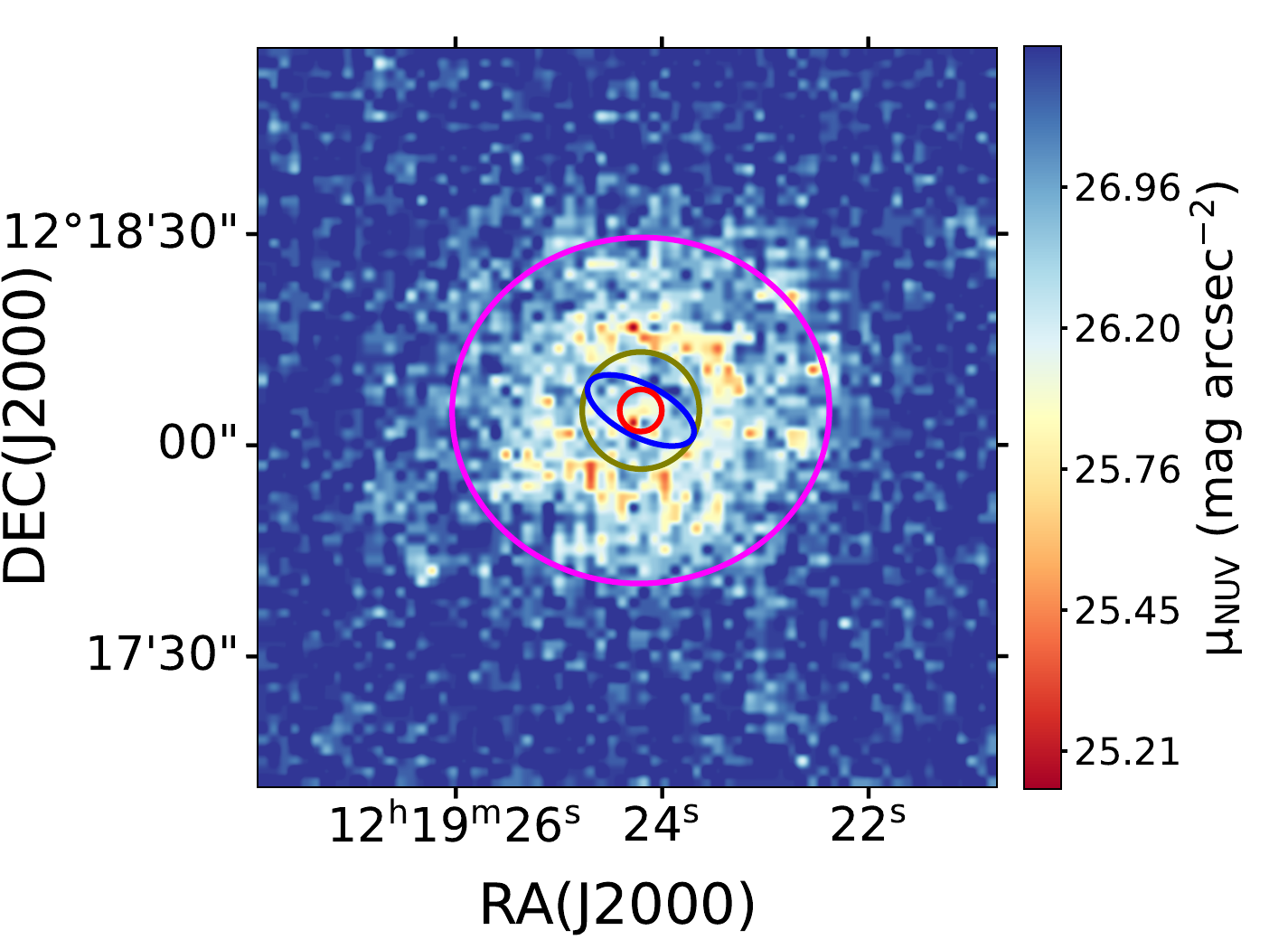}
   \caption{GALEX NUV} \label{fig:x_c}
\end{subfigure}

\bigskip

\begin{subfigure}{0.35\textwidth}
   \includegraphics[width=\linewidth]{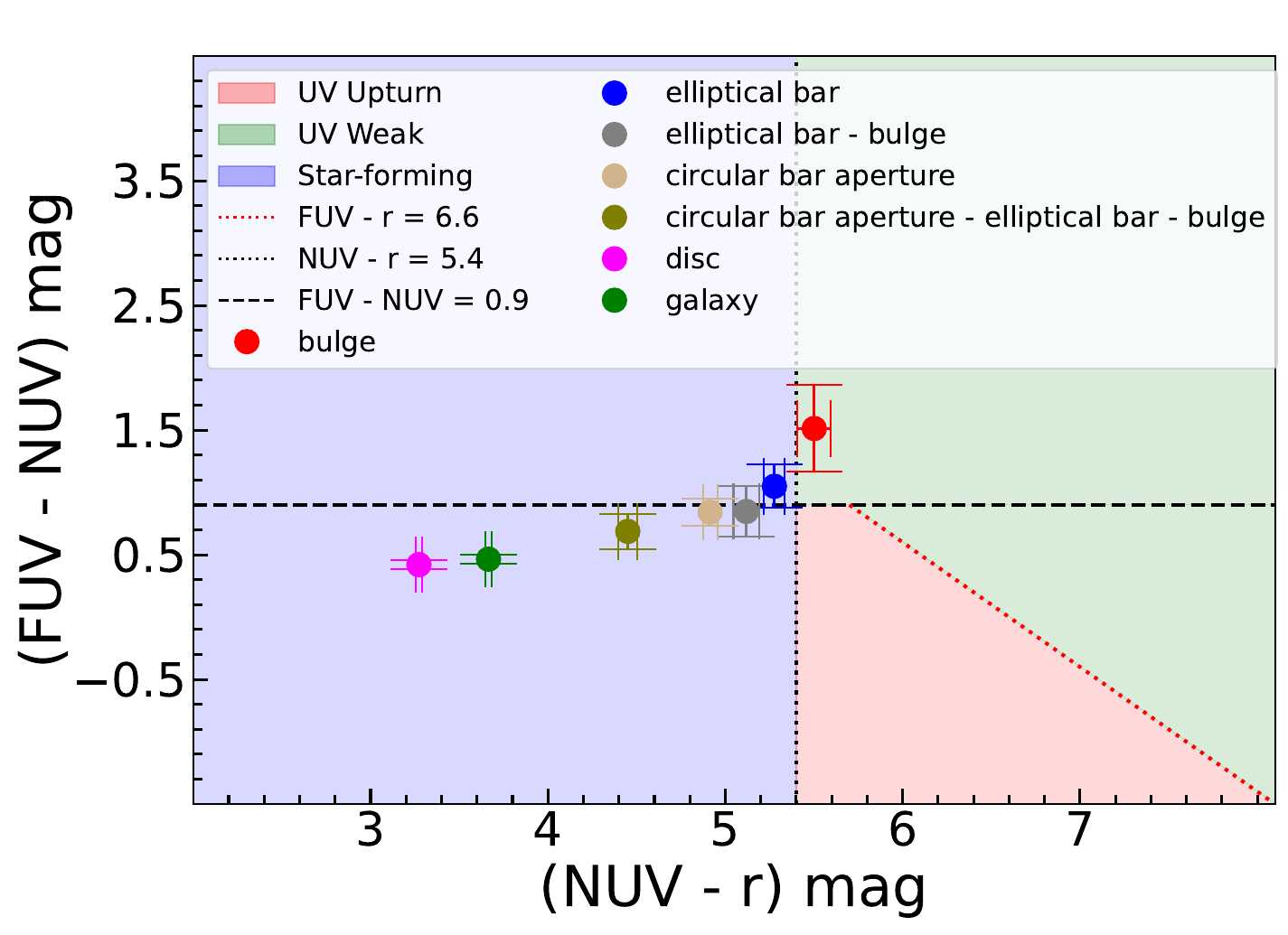}
   \caption{Two Colour Map} \label{fig:x_d}
\end{subfigure}
\begin{subfigure}{0.35\textwidth}
   \includegraphics[width=\linewidth]{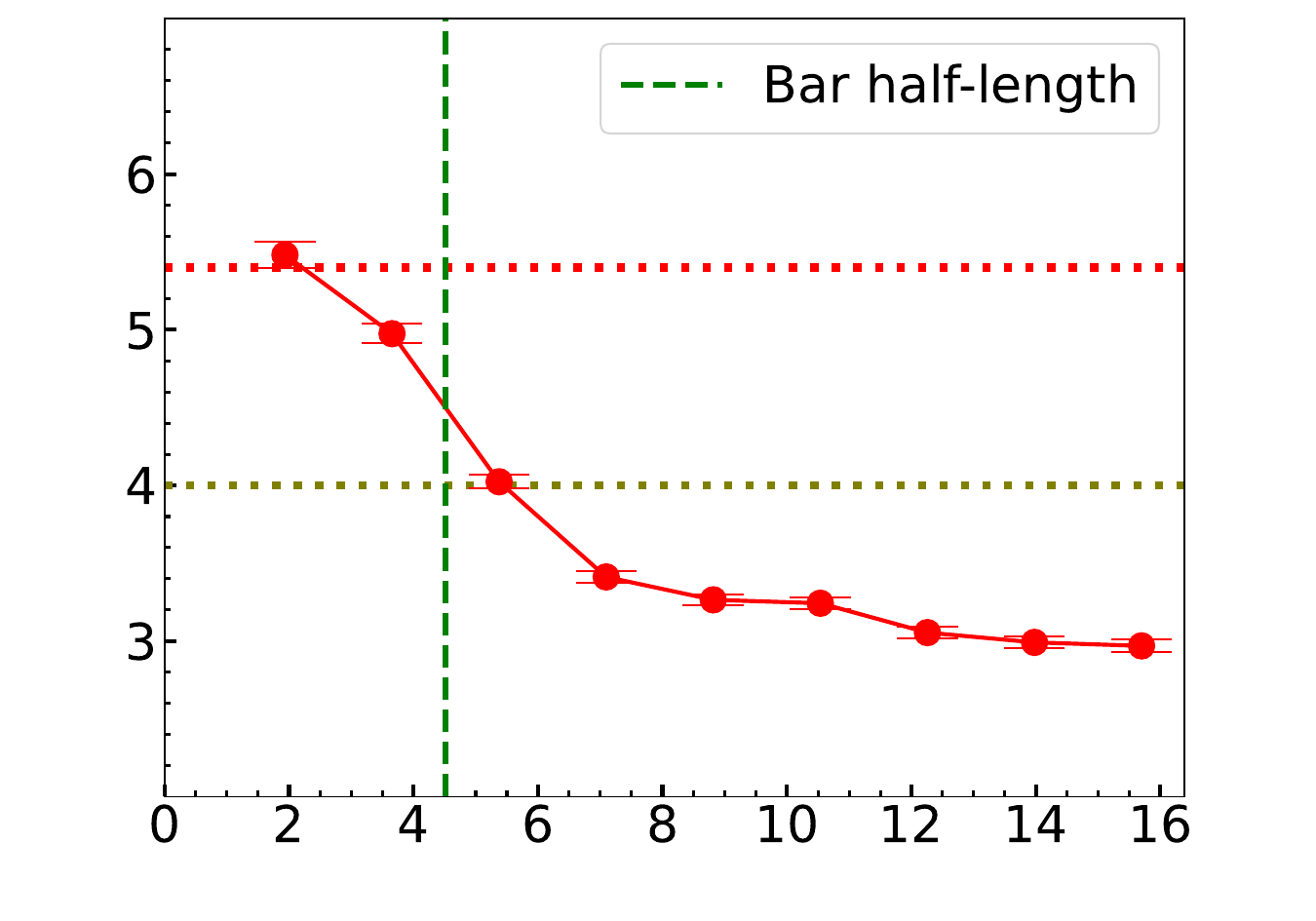}
   \caption{NUV-r radial profile} \label{fig:x_f}
\end{subfigure}
\caption{Steps involved in the analysis (Galaxy no. 6 in Table \ref{table:2}). (a) Four apertures are over plotted on the SDSS r-band degraded image same as in Fig. \ref{fig:4}b, for distinct regions i.e. bulge region (red), elliptical bar region (blue) , circular bar region (brown), and disc (magenta). Outer aperture is defined by R$_{25}$ scale length. The innermost aperture is defined for bulge with fixed diameter of 6 arcsec considering GALEX PSF resolution. The middle circular aperture covers region up to bar length; (b) and (c) Same apertures are over-plotted on GALEX FUV (degraded to NUV resolution) and GALEX NUV image where flux scale is in counts per second. Each of the three band images is provided with a corresponding surface brightness scale for reference. (d) UV-optical spatially resolved colour-colour map for all structural components. The Bulge is lying in UV upturn while the disc is in star-forming region. (e) The NUV$-$r colour radial profile. The red and yellow horizontal dotted lines represent NUV$-$r = 5.4 and 4 mag, respectively. The green vertical dashed line denotes bar half-length. The colours within bar region are redder than the outer disc with the transition is occurring near the end of the bar. }
\label{fig:5}
\end{figure*}

\subsection{Different sub-regions of sample galaxies} 
\label{subsec:subregions}
To obtain the spatially resolved UV-optical colour-colour maps of our 17 sample galaxies, we defined the following regions in the PSF matched and background subtracted, NUV, FUV, and r-band images: circular bulge region (with a diameter corresponding to the PSF of the lowest resolution image, which in our case is the GALEX NUV image and it is $\sim$ 6"), elliptical bar region (defined using the measured bar length, ellipticity and position angle, as described in Sect. \ref{subsec:barlength}), circular bar region (corresponding to a radius equivalent to the semi-major axis of the bar), entire galaxy (defined using an elliptical aperture covering whole galaxy with its major axes defined in terms of $R_{25}$), disc region (which is basically the entire galaxy $-$ circular bar), elliptical bar after subtracting circular bulge region, circular bar after subtracting circular bulge region, and circular bar after subtracting elliptical bar region. In Fig. \ref{fig:4}b, the different apertures defined for bulge (red), elliptical bar (blue), circular bar region (brown), and disc components (magenta) are shown on the original SDSS r-band image of a sample galaxy.
The effective bulge radii of the bulges in our sample galaxies, as given in Table \ref{table:3}, is in the range of r$_{e}$ $\sim$ 0.5 -- 1.1 kpc. In our analysis, we have fixed the bulge size to a radius of 3" and it corresponds to a physical length scale of $\sim$ 1.0 -- 2.5 kpc at the distance of our sample galaxies. We note that by fixing the bulge size to a radius of 3", the defined bulge region might also include parts of the disc or bar regions of the galaxies.
\begin{figure*}[]
    \includegraphics[width=1.0\linewidth]{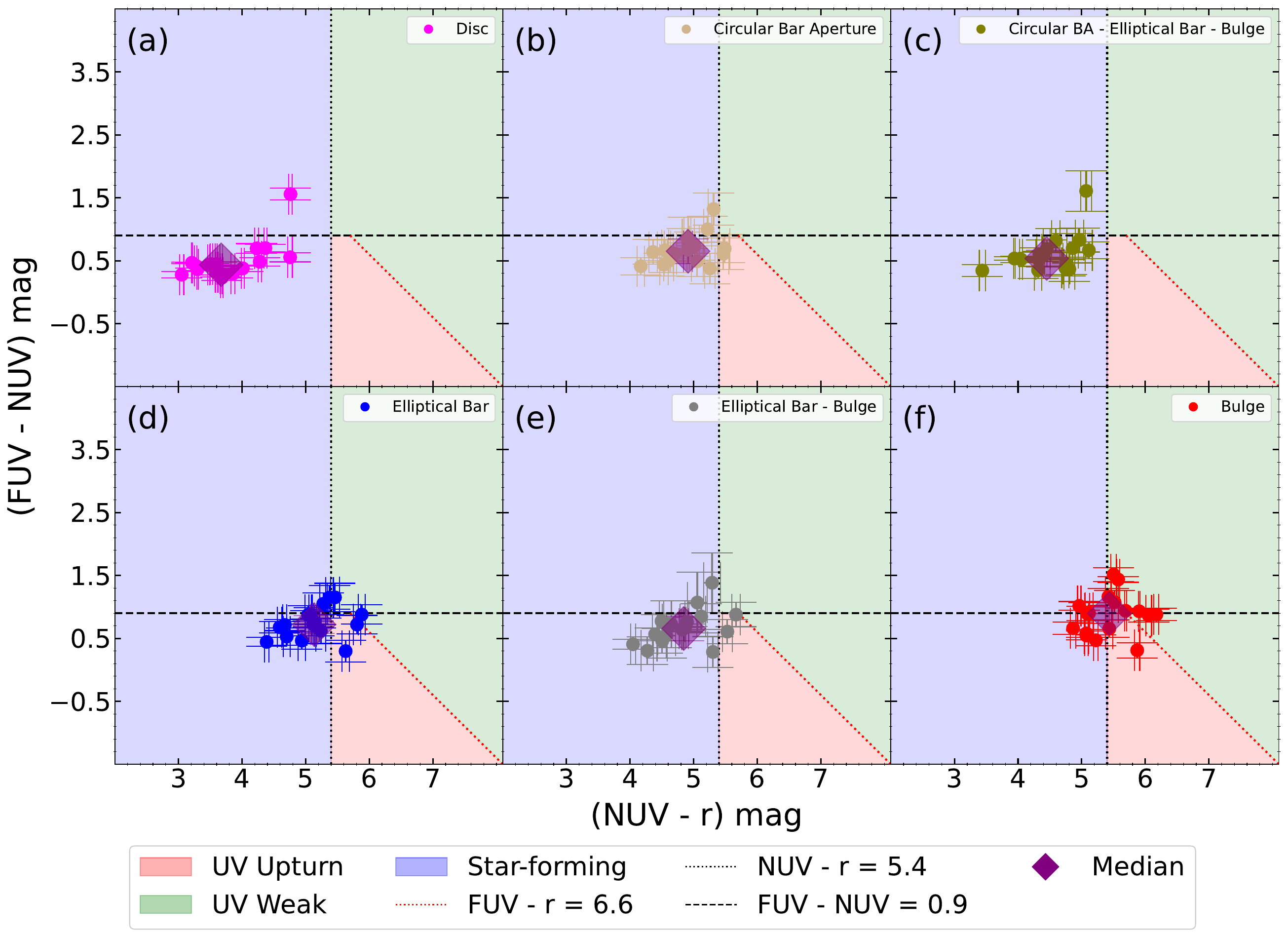}
    \caption{The UV-optical spatially resolved colour-colour maps for (a) the disc component, (b) the bar circular aperture region without subtracting the bulge component, (c) the bar circular aperture after subtracting the bulge and bar components, (e) the bar elliptical aperture region without subtracting the bulge, (f) the bar elliptical aperture region after subtracting the bulge, and (g) the bulge region within 6 arcsecond diameter. The diamond symbol denote the median values in each map. The spatial resolved two colour maps reveal that the disc regions of all galaxies are bluer ($<$ 4 mag) indicating ongoing star formation phase, while inner region, bar and bulge regions of these galaxies within bar corotation radius, are redder ($>$ 4mag) reminiscent of older stellar population with ages $>$ 1Gyr.}
    \label{fig:6}
\end{figure*}
\section{Results}    
\label{sec:result}
\subsection{Spatially resolved UV-optical colour-colour maps}  
\label{subsec:UV2}
To further understand the nature of UV emission in the different regions of our sample galaxies, we analysed the spatially resolved UV-optical colour-colour maps. 
For all the sub-regions (defined in Sect. \ref{subsec:subregions}) of our sample galaxies, the FUV, NUV, and r-band magnitudes are estimated and corrected for the foreground extinction using the values taken from \cite{2011ApJ...737..103S}. From the extinction-corrected magnitudes, we derived the colours and created UV-optical colour-colour maps, which help us to understand the nature of star formation in these sub-regions. Each component (e.g., bulge, bar, and disc) of the galaxy can be identified to be star-forming or quenched depending upon its location in the two colour map. 
The steps involved in the analysis of a sample galaxy (galaxy no. 6) are shown in Fig. \ref{fig:5}. The different apertures as previously described, defined for bulge, bar, and disc are over-plotted on the resolution-matched SDSS r-band, GALEX FUV, and GALEX NUV images and are shown in Figs. \ref{fig:5}a-\ref{fig:5}c.
The positions of the different sub-regions in the two colour map are shown in Fig. \ref{fig:5}d. In this sub-fig., the bulge (red point) lies in UV weak (NUV$-$r $>$ 5.4 mag and FUV$-$NUV $>$ 0.9 mag) region indicating a lack of strong UV emission. The bar region, with and without subtracting bulge (represented by grey and blue points respectively) component, exhibit a NUV$-$r colour more than 5 mag. Similarly, the circular bar region (brown point) after subtracting both bulge and elliptical bar component display colour greater than 4.5 mag which translates to stellar population older than 1 Gyr (explained in next Sect.). This implies no star formation inside the bar region, in the last 1 Gyr. This is in contrast to the disc region (magenta point) which displays much bluer NUV$-$r colour ($\sim$ 3.3 mag) suggesting the presence of ongoing star formation in the last 1 Gyr. The redder values of NUV$-$r colour up to circular bar region, indicates that the inner regions, up to the bar length, are quenched. The colours of the disc suggest that the disc is actively star-forming.\\
In Fig. \ref{fig:6}, we present spatially resolved UV-optical colour-colour map for the entire sample of 17 galaxies. The median value for each sub-region across the 17 galaxies is also shown in sub-plots. The disc regions of these galaxies are consistently bluer than their inner structural counterparts while the circular bar regions with or without subtracting bulge and bar component, exhibit redder NUV$-$r colours than the disc regions, typically greater than 4 mag. The elliptical bar regions with or without subtracting the bulge component are even more redder. The bulges mostly fall in UV weak or UV upturn, or closer, indicating that the detected UV emission is predominantly coming from older stars, not from young stars. This is more in line with inside-out quenching scenario happening in galaxies. In the next Sect., we attempt to quantify the age of the stellar populations in these different regions using the NUV$-$r colour as proxy for age.
\begin{figure*}[]
    \includegraphics[width=1.0\linewidth]{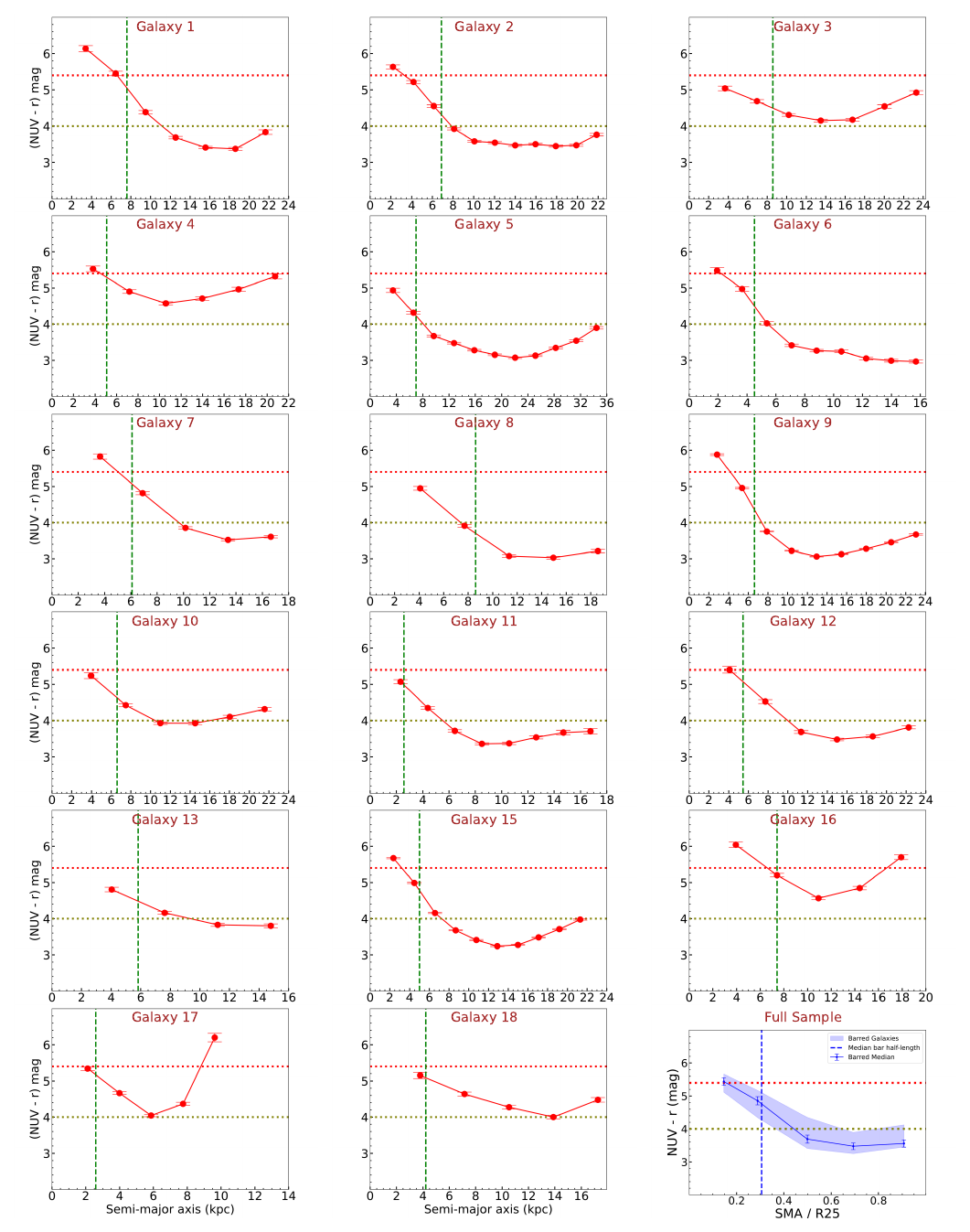}\
    \caption{The NUV-r radial profile is shown for all 17 galaxies, with the red and yellow dashed horizontal lines reference NUV-r = 5.4 and NUV-r = 4 and the vertical line representing bar half-length. The bottom-right most sub-figure presents median NUV-r values with error bars plotted against normalised SMA/R$_{25}$ for the full sample, with the shadow region representing the inter quartile range (IQR) of the y-values within each bin, bounded by the 25th and 75th percentiles. The dashed blue line indicates the median normalised bar half-length. The results are discussed in Sect. \ref{sec:result}. }
    \label{fig:7}
\end{figure*}

\subsection{NUV$-$r profile}
\label{subsec:nuv}
\begin{figure}
\centering
\includegraphics[width=\linewidth]{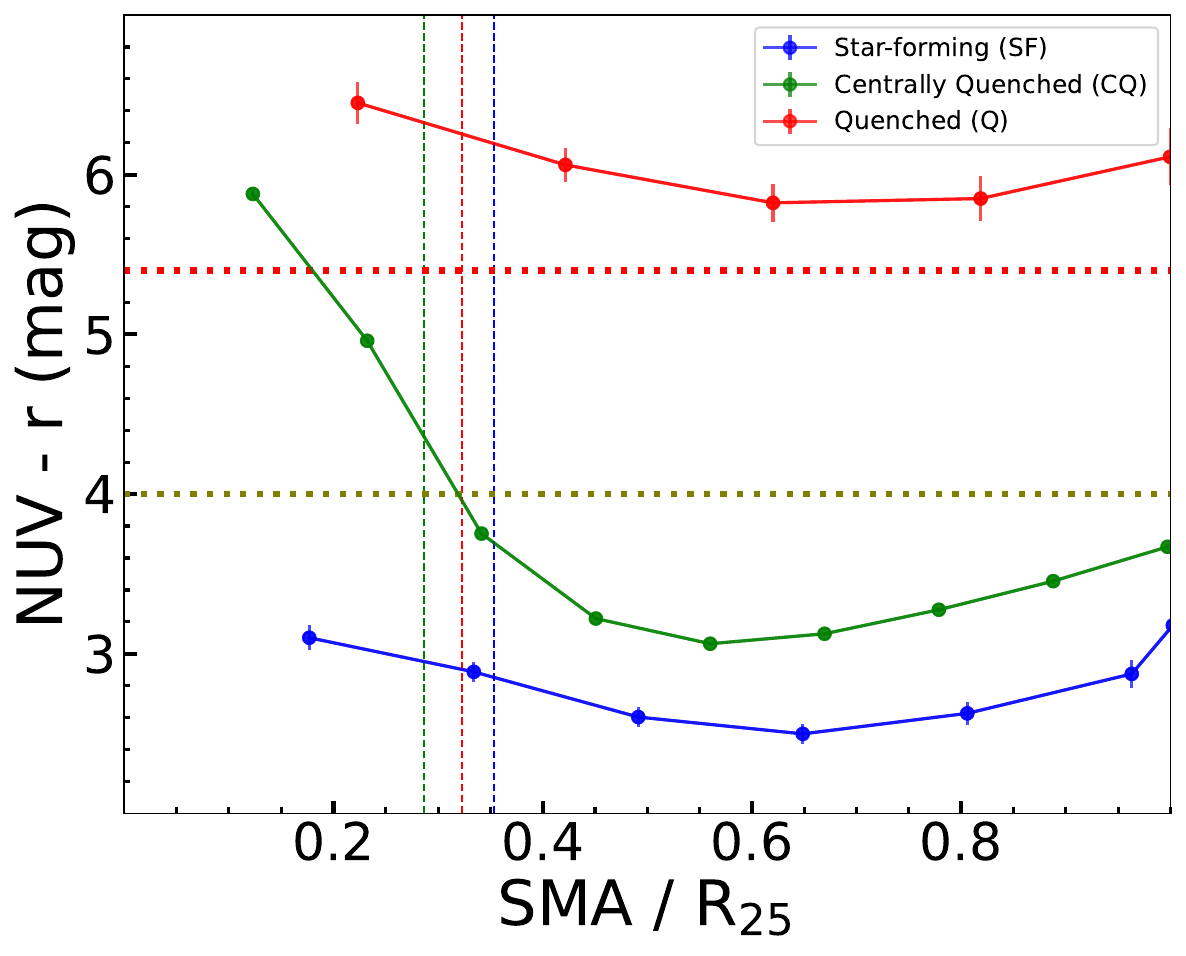}
\caption{The NUV-r normalised radial profile for three barred galaxies positioned differently on the SFR-M${\ast}$ plane is shown with corresponding colour-coded bar half-lengths. The star-forming barred galaxy (blue) lies in the star-forming region, as determined by both the MPA-JHU \citep{brinchmann2004physical} and GSWLC \citep{2018ApJ...859...11S} catalogues. The quenched barred galaxy (red) is positioned in the quenched region in both catalogues, while the centrally quenched galaxy (Galaxy 9) is selected from our sample. All three galaxies share similar stellar masses around the median value (logM${\ast} \sim 10.9$ M$_{\odot}$), within comparable redshifts and axial ratios (b/a). The figure clearly pictures the role of bars in driving barred galaxies from the MS to the quenched region by inducing a cessation of star formation.}
\label{fig:8}

\end{figure}

Massive O and B type stars in recent star-forming regions (up to 300 Myr), emit in UV and hence the UV emission arising from the galaxies can be used as a direct tracer of recent star formation \citep{2012ARA&A..50..531K}. However, the FUV flux might be contaminated by the flux from the old hot stellar population. NUV is less contaminated by the emission from old population. Based on the analysis of single stellar population models with different metallicities, \citet{Kaviraj_2007} showed the evolution of NUV$-$r as a function of age. They showed that the NUV$-$r colour evolves slowly after 1 - 2 Gyrs (Fig. 7 in \citet{Kaviraj_2007}) for all models and an intrinsic NUV$-$r colour of 4 mag corresponds to $\sim$ 1 Gyr or higher. This is because the young stars contributing to the NUV flux evolves off the MS in this timescale. They suggested that NUV$-$r colour is an effective age indicator for the population formed in the last $\sim$ 1 Gyr. Therefore, the NUV$-$r colour can be used to determine whether a star formation event has occurred within the past 1 Gyr. \cite{Pan_2016} found that for less dust-attenuated galaxies, NUV$-$r is found to be linearly correlated with D$_n$4000. D$_n$4000 index is an excellent age indicator and it represents spectral discontinuty around 4000 $\AA$ wavelengths. This happens due to either the absence of young UV-dominated stars or increased metal absorption in the stellar atmosphere of older metal-rich stars. \cite{Pan_2016} found that NUV$-$r colour of $\sim$ 4.0 mag corresponds to a D$_n$4000 value of 1.5. A value of $<$ 1.5 indicates the presence of stellar population younger than 1 Gyr and a value $>$ 1.5 indicates a population older than 1 Gyr \citep{kauffmann2003stellar}. Thus, NUV$-$r is a good photometric indicator of stellar age. This is similar to the correlation found by \cite{2014SerAJ.189....1S} between the NUV$-$r and the specific SFR. \\ 
As shown in Fig. \ref{fig:6}, the median NUV$-$r colour of all the regions within a circular aperture equivalent to the bar length is $>$ 4 mag. This indicates that the stellar populations inside the bar region are older than 1 Gyr or there was no star formation in the inner regions (covering the bar) of these galaxies in the last 1 Gyr. However, the median NUV$-$r colour of the disc regions of our sample is $<$ 4 mag (magenta diamond symbol in the upper-left panel of Fig. 6) and it indicates star formation, in the last 1 Gyr, in the outer disc of most of our sample galaxies. In order to better understand this inner to outer transition, we plotted the NUV$-$r colour radial profiles of all the 17 galaxies and shown in Fig. \ref{fig:7}. To create NUV$-$r colour radial profile, we plotted concentric elliptical annuli with a width of 3 arcseconds each, extending out to the R$_{25}$ radius on both SDSS r-band (degraded to the GALEX NUV spatial resolution) and GALEX NUV images, and then estimated the NUV$-$r colour in each annulus and plotted it as a function of the semi-major axis (in kpc). As can be seen from Fig. \ref{fig:5}e, the NUV$-$r colour changes from red to blue as a function of radius and it happens near bar radius. Similarly, in most of the galaxies (Fig. \ref{fig:7}), the transition to a colour $\leq$ 4 -- 4.5 mag happens at distances of $\sim$ 5 -- 9 kpc from the centre of the galaxies, which is similar to the length of the semi-major axis of the bars in these galaxies (shown as green vertical lines in Fig. \ref{fig:7}). 
This shows that inside the bar region, there is no star formation over a period $>$ 1 Gyr and outside the bar there is significant star formation. This timescale of $\sim$ 1 Gyr matches with the results by \cite{2016MNRAS.457..917J}, who found that the star formation between the central sub-kpc region and the ends of the bar in four nearby galaxies to be truncated at least 1 or more Gyrs ago. Our results suggests that these galaxies are centrally quenched and no recent star formation in the last 1 Gyr or more, and the size of this centrally quenched region correlates with the size of the bar. This indirectly indicates the role of bar in central quenching. 
We note that for certain galaxies in our sample, specifically galaxy nos. 3, 4, 16, 17, and 18, the NUV$-$r colour profiles do not exhibit the transitions to bluer colours and remain $>$ 4 mag throughout their extent, up to the R$_{25}$. For galaxies 16 and 17, this is primarily attributed to contamination from foreground or background objects within their fields. Though, we have masked these objects, the contamination still contributes to their profiles. The galaxy 3 has a higher inclination and a higher redshift, resulting in fewer data points in its colour profile, limiting the analysis. However, galaxies 4 and 18, despite sharing similar parameters with the rest of the sample, do not exhibit the typical transition from redder to bluer colours, unlike most galaxies. These galaxies require further investigation.
In the last panel of Fig. \ref{fig:7}, we have shown the median NUV$-$r colour for full sample of 17 barred galaxies as a function of radius normalised with R$_{25}$. Each point represents the median NUV$-$r colour within a bin size of 0.2 and the shadow region represents the inter-quartile range for y-values within each bin, bounded by the 25th and 75th percentiles. The dashed blue vertical line indicates the median bar radius or bar half-length. The average colour transition is happening around 4.5 mag indicating the bar region of most of the barred galaxies is quenched. \\ 
To understand the NUV$-$r colour profiles of barred galaxies in different stages of star formation, with respect to their positions on the M$_\ast$ - SFR plane, and to compare with our sample of centrally quenched galaxies, we selected two representative galaxies: a star-forming barred galaxy located in the star-forming region and a fully quenched barred galaxy located in the quenched region in the M$_\ast$ - SFR plane, based on the parameters from both the MPA-JHU \citep{brinchmann2004physical} and GSWLC \citep{2018ApJ...859...11S} catalogues. Both these galaxies have stellar masses around the median stellar mass value of our sample, log(M$_\ast) \sim 10.9$ M$_\odot$, comparable redshifts and axial ratios (b/a) as those of our sample galaxies and do not host AGN. Their normalised bar lengths are similar to that of the median bar length of our sample galaxies. The two galaxies also share similar bulge properties to that of our sample galaxies (s\'ersic indices, n$<$2). 
We compared the NUV$-$r normalised colour profiles of these two galaxies with those of our sample of centrally quenched galaxies. Fig. \ref{fig:8} shows the typical NUV$-$r profiles of these star-forming barred galaxy and fully quenched barred galaxy, along with the profile of one of our sample galaxies with similar stellar mass.\\
The inner region of the star-forming galaxy exhibits colours well below 4 mag, with a nearly flat NUV$-$r profile, with its disc slightly bluer than the inner region. This suggests that the bar is still actively funneling gas and driving star formation in both the central regions and across the bar. In contrast, the (NUV$-$r) profile of the quenched galaxy shows colours exceeding 5.4 mag throughout its disc, indicating it is fully quenched, with the disc lacking the fuel needed for star formation. Fig \ref{fig:8}. illustrates this progression, with MS galaxy exhibiting ongoing star formation throughout the galaxy, while quenched galaxy is fully quenched at all radii. Our sample of barred galaxies lies in between, showing central quenching up to the extent of the bar but retaining star-forming discs. This might indicate an evolutionary sequence of a barred galaxy moving from a star-forming stage to a quenched state, mainly by the action of bar. Although the potential impact of strong bars on disc properties requires further exploration, comparing our results with galaxies at different stages of star formation suggests that bars play a significant role in suppressing star formation in the central regions of internally quenched galaxies. A more detailed analysis of a larger sample of galaxies in different evolutionary stages, based on their location in the M$_{\star}$ - SFR plane, will be presented in a future work.

\begin{figure*}
\centering
\begin{subfigure}{0.4\textwidth}
   \includegraphics[width=\linewidth]{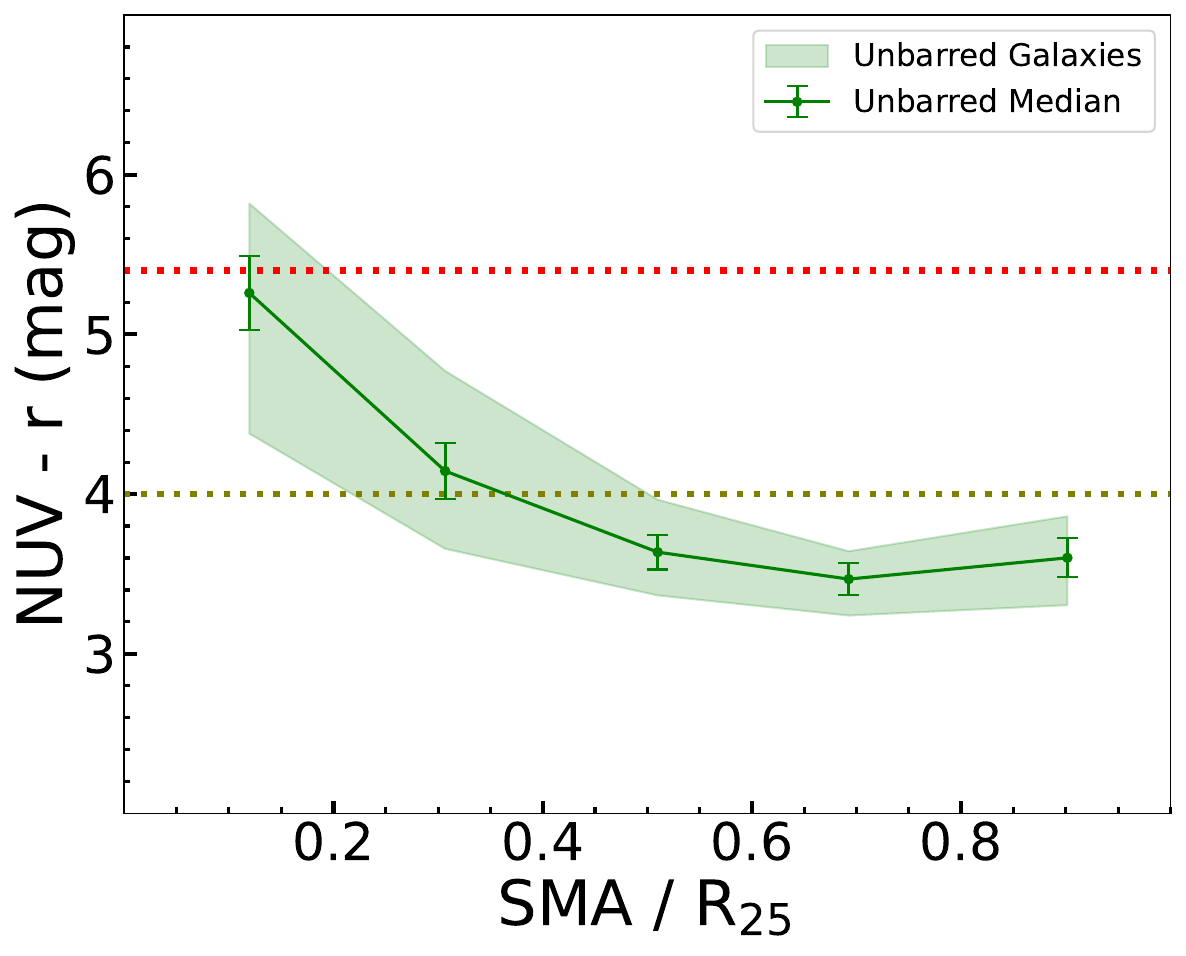}
   \caption{} \label{fig:x_a}
\end{subfigure}
\begin{subfigure}{0.4\textwidth}
   \includegraphics[width=\linewidth]{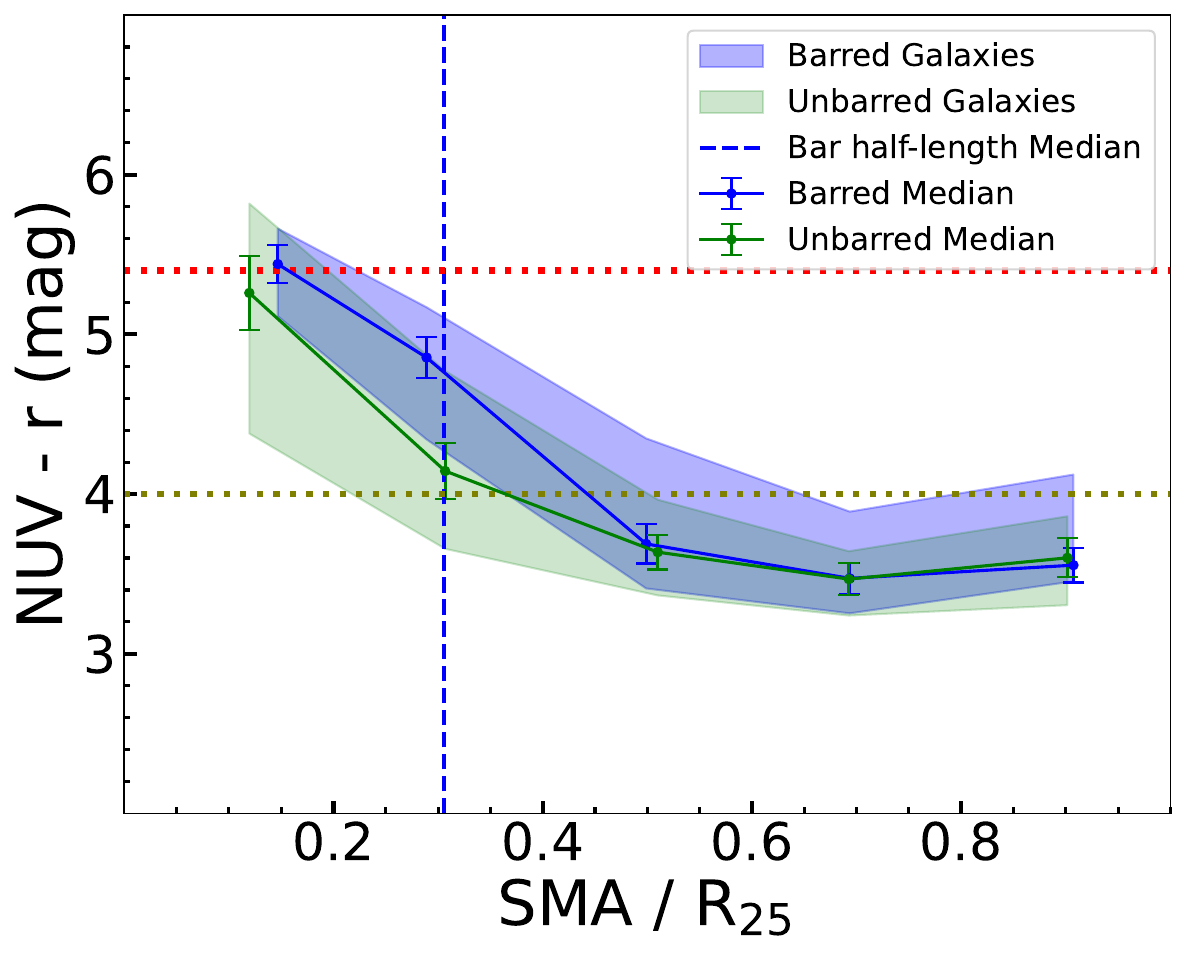}
   \caption{} \label{fig:x_a}
\end{subfigure}
\begin{subfigure}{0.4\textwidth}
   \includegraphics[width=\linewidth]{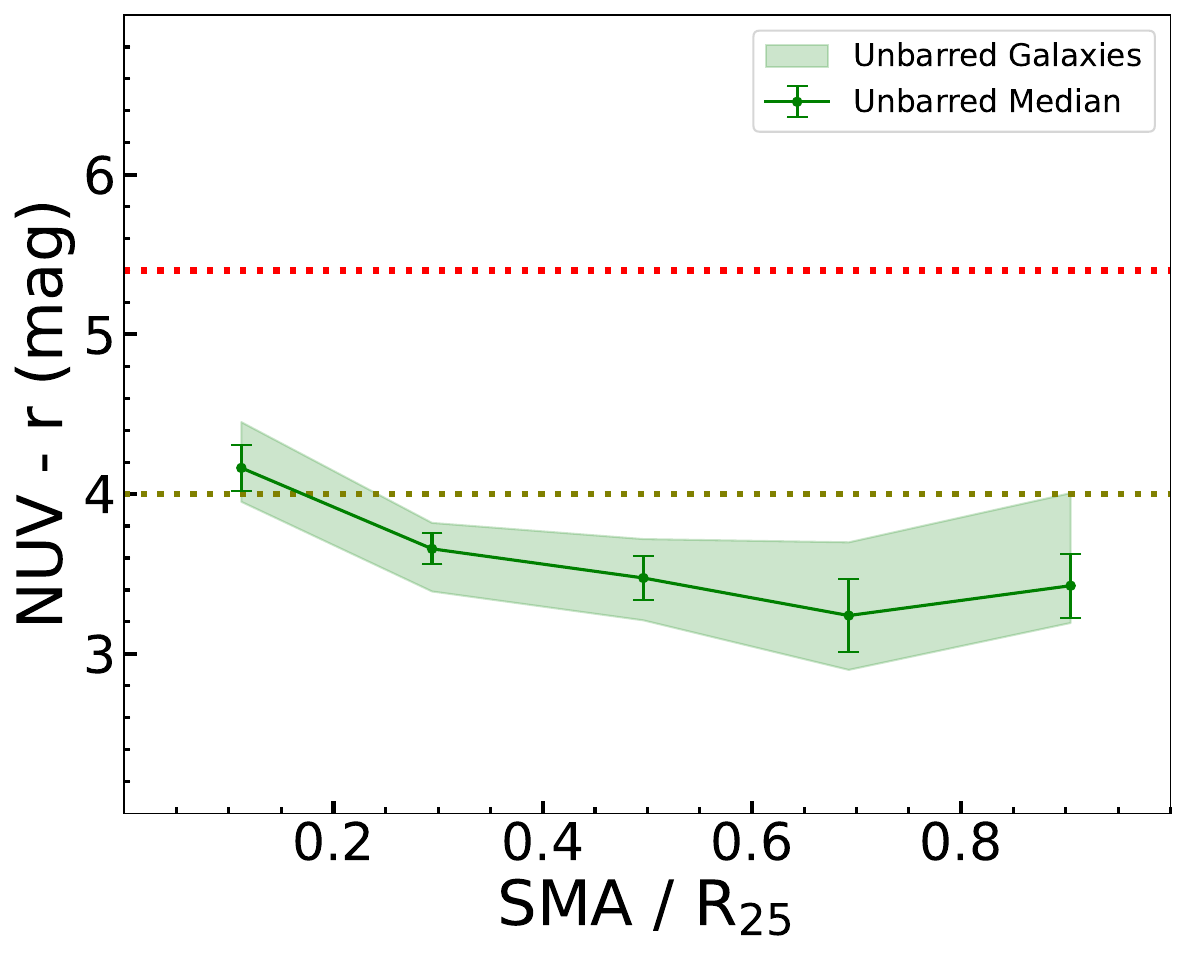}
   \caption{} \label{fig:x_a}
\end{subfigure}
\begin{subfigure}{0.4\textwidth}
   \includegraphics[width=\linewidth]{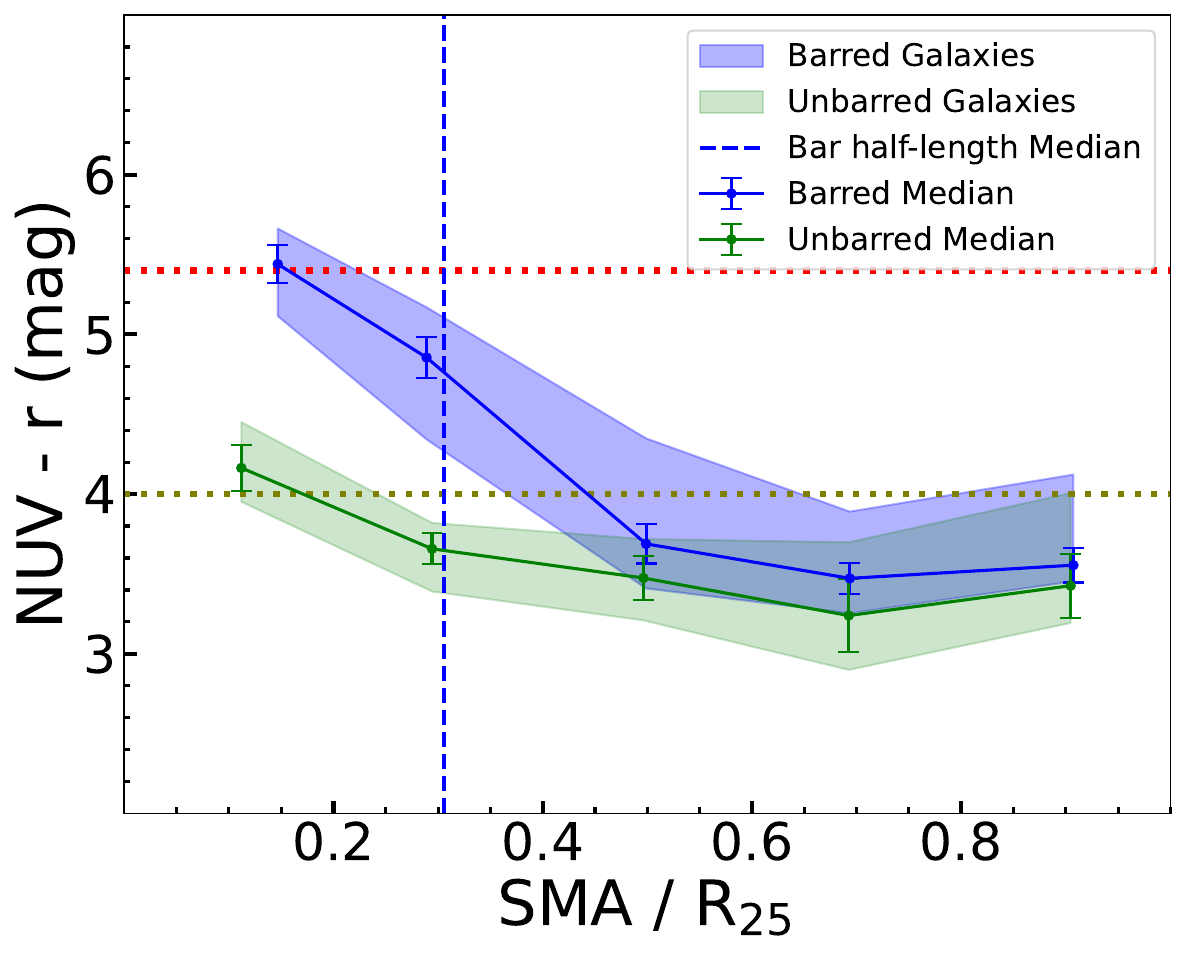}
   \caption{} \label{fig:x_a}
\end{subfigure}
\caption{(a) The median NUV$-$r values are plotted against the normalised SMA/R$_{25}$ for the control sample of 8 unbarred galaxies. (b) The median NUV$-$r profiles for the complete sample of barred and unbarred galaxies.(c) The median profile for unbarred galaxies after excluding four internally quenched unbarred galaxies. (d) The profiles of both barred and unbarred sample, after removing the four internally quenched unbarred galaxies. The results show that the inner regions of barred galaxies are quenched within their bar length, whereas unbarred galaxies exhibit comparatively bluer profiles.}
\label{fig:9}
\end{figure*}

\section{Discussion}
\label{sec:dis}
In this study, we used the spatially resolved UV-optical colour-colour maps and NUV$-$r radial profiles of 17 centrally quenched barred galaxies, in the redshift range of 0.02 -- 0.06, to understand the role of bars in quenching star formation.  
We utilized spatially resolved (FUV$-$NUV vs. NUV$-$r) colour-colour maps to understand the nature of UV emission in different sub-regions of these galaxies. We found that our sample galaxies exhibit notably redder colours, within their bulge, bar and the region within the co-rotation radius of the bar, suggesting UV emission from old and hot stellar population in the bulge region and lack of recent star formation in the bar region, while their discs remain bluer, reflecting ongoing star formation. Then to better quantify the age of the stellar populations, we used NUV$-$r colour as a proxy of age and created the NUV$-$r radial profiles of these systems. These profiles suggested that the NUV$-$r colour changes from red to blue as a function of radius and the stellar population inside the bar co-rotation radius has ages $>$ 1 Gyr. The disc region of majority of our sample galaxies (12 out of 17) have NUV$-$r $<$ 4 mag, indicating recent star formation (in $<$ 1 Gyr) in the discs. The colour transition from red to blue (at NUV$-$r $\sim$ 4 mag) is particularly evident at the bar radius, emphasizing the bar's role in suppressing star formation. Strong and extended bars are known to inhibit star formation by either funneling gas to the centre or producing shear in the gas that prevents its collapse. The HI gas holes are observed in case of removal of gas when the period of intense star formation is over after the bar has completely funneled the gas all the way to the centre \citep{george2020more, gavazzi2015role, 2020MNRAS.492.4697N}. Such HI cavities are observed in very nearby galaxies, suggesting that the redistribution of gas by bar is the mechanism of bar-quenching in these galaxies \citep{ george2019insights, george2020more}. The HI interferometric observations of these barred spiral galaxies to know the presence or absence of HI gas in the bar region will be of interest to understand the physical mechanisms of bar quenching in these redshifts. \\
We note that in our analysis we have only considered galactic foreground extinction correction and do not apply any internal extinction correction to the NUV, FUV, and r-band magnitudes. This suggests that the estimated colours of different regions of the galaxies are an upper limit and hence the related ages of these regions. To check the effect of internal extinction in the estimated colours, we used the dust attenuation value for each of our sample galaxies, provided in the GSWLC M2 catalogue, and re-estimated the colours of different regions of our sample galaxies. The dust attenuation in the GSWLC catalogue is determined by constraining SED fits with infrared luminosity (SED+LIR) and represents the internal extinction. We used a constant attenuation value across the galaxy. As expected, the re-estimated colours are bluer. But the change in colours of different regions, with and without internal dust attenuation, is of the order of $\sim$ 0.25 mag. Even with this shift in colour to a bluer value, the (NUV$-$r) colour of our sample galaxies, inside the bar co-rotation radius remains to be $>$ 4 mag and hence, the age of the stellar population can be assumed to be older than 1 Gyr.  As we are using a constant value of dust attenuation, the shape of the NUV$-$r profiles are not affected. However, the dust attenuation can be a function of galacto-centric radius and such variation can affect the NUV$-$r profile of our galaxies. The studies of two-dimensional dust maps of nearby galaxies and those at z $\sim$ 1.4  by \cite{2015A&A...581A.103G} and \cite{2016ApJ...817L...9N} respectively, indicate that in the galaxy scale, the dust attenuation is not a strong function of radius. These studies suggest that, for galaxies massive than 10$^{10}$ M$_{\odot}$, the dust attenuation could have a strong negative gradient in the inner 1 kpc region, but beyond that the dust attenuation has no variation. As the spatial resolution of the NUV$-$r profile of all our sample galaxies are coarser than 1 kpc, we do not expect a change in their profile due to the variation in the dust attenuation across the galaxy. Thus, the effect of internal extinction in the estimated colours and the order of magnitude estimation of the ages of the stellar populations in the different regions of our sample galaxies is likely to be negligible and our results about the role of bars in quenching star formation in the central regions are expected to be unaffected. \\
The spatially resolved UV-optical colour-colour maps and the  NUV$-$r colour profiles of our sample galaxies (with most of them having stellar mass greater than 10$^{10.5}$ M$_\odot$) suggest an inside-out quenching scenario for these galaxies. Such an inside-out quenching scenraio is expected for high mass ($\ge$ 10$^{10.5}$ M$_\odot$) galaxies (\citealt{2015ApJ...804L..42P,2019ApJ...872...50L,2019ApJ...884L..52Z,2021ApJ...911...57Z}).  
However, these previous studies suggest different physical mechanisms, such as morphological quenching, bar-quenching and AGN feedback, for the inside-out quenching in secularly evolving high mass galaxies. \cite{2019ApJ...872...50L}  favoured morphological quenching over AGN feedback, as the primary mechanism. Morphological quenching \citep{2009ApJ...707..250M} is  mainly driven by classical bulges. The S\'ersic indices of the bulges, given in Table \ref{table:3}, of our sample galaxies are  $<$ 2 (\citealt{2008AJ....136..773F,2010ApJ...716..942F}), indicating the pseudo nature of these bulges. Hence, the quenching of central regions of our sample galaxies due to the action of bulges is highly unlikely. As described earlier, none of our sample galaxies host AGN. However, a past AGN activity causing central quenching of these systems cannot be ruled out. But again, it is highly unlikely that all our sample galaxies to have a history of past AGN activity which created a centrally quenched region exactly similar to the region covering the co-rotation radius of the bar.  \cite{2020MNRAS.495.4158F} suggested that the physical properties of a bar are mostly governed by the stellar mass of the host galaxy and found that massive galaxies host longer bars. \cite{george2021bar} observed a mild correlation between the offset of fully quenched barred galaxies from the SFMS, and the scaled bar length for isolated massive galaxies.  These results indicate a role of bar in the inside-out quenching of high mass galaxies. Hence, the most likely mechanism for the observed inside-out quenching in our sample galaxies is the action of bar.\\
To further validate our findings and determine if the central quenching is only associated with the bars, we conducted a comparative analysis with a control sample of 8 unbarred galaxies, as shown in Fig. \ref{fig:9}. These unbarred galaxies are also centrally quenched and are selected using the same criteria outlined in Sect. \ref{sec:SS}. We also note that these unbarred galaxies dust attenuation levels are similar with our main sample of barred galaxies. The properties of this control sample of unbarred galaxies are mentioned in Table \ref{table:4}.  We obtained the individual NUV$-$r profiles of all these 8 unbarred galaxies and then obtained a median profile for our sample. We plotted the NUV$-$r colour against the normalised disc radius, with each data point representing the median NUV$-$r values within a bin size of 0.2 for both barred and unbarred sample galaxies. The median bar radius of our barred sample galaxies is also indicated as blue vertical line in Figs. \ref{fig:9}b and \ref{fig:9}d. Although the NUV$-$r profiles appear similar for both barred and unbarred samples in Fig. \ref{fig:9}b, the unbarred galaxies exhibit a larger variance in their colour distribution (Fig. \ref{fig:9}a). When the individual profiles of the unbarred galaxies are checked, we found that this increased variance is primarily due to the presence of four unbarred galaxies with quenched central regions. To understand the cause of this central quenching in these unbarred galaxies, we performed the two-dimensional structural analysis using GALFIT \citep{2010AJ....139.2097P} to infer the properties of bulges in our sample of unbarred galaxies sample. The estimated parameters are given in Table \ref{table:4}. Our analysis reveals a dual nature for these bulges in the unbarred galaxies. Five of the unbarred galaxies, have their bulge Sérsic indices $>$ 2, indicating that they are classical in nature (\citealt{2008AJ....136..773F,2010ApJ...716..942F}). Four of these five galaxies (galaxy no. 2, 4, 5, and 8) are those which have their central regions quenched to a larger extent, with their NUV$-$r radial profiles similar to the profiles of barred galaxies. This suggests that the central quenching in these four unbarred galaxies, to a larger radial extent, similar to extent that is observed in the barred galaxies, could be due to the action of classical bulges.  Interestingly, we found that the galaxy no. 4 in Table \ref{table:4}, showed some indication for the presence of a bar like feature in the residue of our GALFIT analysis. But an addition of a bar component did not provide a satisfactory fit, probably due the higher redshift of this sample galaxy and/or lower b/a value. Thus, for this galaxy, central quenching due to the action of bar cannot be ruled out. For a viable comparison with the barred galaxies, we removed the four centrally quenched unbarred (galaxy no. 2, 4, 5, and 8) galaxies which show different bulge properties than those of our barred galaxies sample. The remaining four systems show redder colour values NUV$-$r $>$ 4 only to a smaller extent from the centre, while bluer profiles with NUV$-$r $<$ 4 mag (Fig. \ref{fig:9}c) are observed for the rest of the region. While comparing our barred sample with these four systems (Fig. \ref{fig:9}d), it becomes evident that barred galaxies display redder internal colours than unbarred galaxies, a trend that persists up to the extent of the bar. We here note that the limited size of our barred galaxy sample and the control sample of unbarred galaxies may have implications when generalising the observed results to a larger sample. Since this was a pilot study aimed at validating the methodology, we plan to confirm our findings with a larger sample in the future work. \\
As detailed in Sect. \ref{sec:SS}, we selected our sample by leveraging the discrepancies between two catalogues, which allowed us to identify centrally quenched galaxies. In the final selection step, we excluded emission-type galaxies—specifically, those exhibiting H$\alpha$ emission in their SDSS optical spectra within the 3-arcsecond fiber aperture. The presence of H$\alpha$ emission might indicate ongoing star formation in the sub-kpc nuclear regions, which we avoided in our analysis. This was mainly to focus on those galaxies where star formation in the central regions has completely ceased. However, it would be interesting to study these emission-type galaxies, as they can trace the final episodes of star formation. By probing the inner regions with deeper and higher-resolution data, we could gain valuable insights into the processes driving star formation in these nuclear regions just before quenching occurs. Such a study will be of interest to understand different stages of bar-quenching, as observed for a small number of nearby galaxies in \cite{george2020more}. Moreover, to further facilitate our findings, we plan to employ SED fitting to map the star formation history and the ages of stellar population in centrally quenched galaxies, where bars play a crucial role in regulating star formation. Spatially resolved stellar population age maps and star formation history maps of these centrally quenched barred galaxies will also provide insights to understand the role of bars in radial migration of stars (\citealt{2019MNRAS.489.4992D,2024MNRAS.534.2438N}).\\
In this study we used the spatially resolved colour-colour maps to analyse a small sample of barred galaxies in the redshift range of 0.02 - 0.06. As stated before, we plan to apply this method to a larger sample of SDSS galaxies classified using Galaxy Zoo. The sample will be further refined to include barred galaxies at different stages of evolution, based on their position in the SFR-M$_{\ast}$ plane.
   
\section{Summary}
\label{sec:summary}
In this pilot study designed to investigate the role of bar in quenching star formation in barred galaxies, we selected a sample of nearly face-on centrally quenched barred galaxies, within a redshift of 0.02 -- 0.06, and studied them using spatially resolved UV-optical colour-colour maps. We also analysed their NUV$-$r radial profiles to understand the age of stellar populations in these regions. 
\begin{itemize}
    \item A sample of 17 barred galaxies are identified as centrally quenched galaxies, based on their location in the SFR vs stellar mass (SFR - M$_{\ast}$) plot created using the parameters from the MPA - JHU and GSWLC catalogues.
    \item Spatially resolved UV-optical colour-colour maps of these galaxies are used to understand the nature of UV emission in different regions of galaxies. The location of different sub-regions in the colour-colour maps and their median colours, indicate UV emission from old and hot stellar population in the bulge region and lack of recent star formation in the bar region, while presence of recent star formation in the disc region.
    \item All the centrally quenched barred galaxies display redder colours (NUV$-$r $>$ 4 mag) in the inner regions, up to the length of the bar, indicating the age of stellar population in these regions to be older than $>$1 Gyr. This suggests the role of the bar in suppressing star formation in these galaxies. 
    \item Most of our sample galaxies (12 out of 17 galaxies) also exhibit a transition from red to blue in the NUV$-$r colour profile  (NUV$-$r $<$ 4 mag) when moving from the inner region to the disc region, indicating star formation activity in the disc region in the last 1 Gyr.
    \item Most barred galaxies in our sample host pseudo bulges and do not host AGN, indicating that the most probable reason for the internal quenching of these galaxies is the action of stellar bar. 
    \item The barred galaxies show redder colours (NUV$-$r $>$ 4 mag) or host older population ($>$ 1 Gyr) to a larger spatial extent compared to their unbarred counterparts, lying in a similar regime of stellar mass and redshifts. 
\end{itemize}

\begin{acknowledgements}
The authors thank the anonymous referee for the constructive comments which have helped to improve the manuscript. This study made use of archival Sloan Digital Sky Survey (SDSS) Data Release 7 (DR7) and Galaxy Evolution Explorer (GALEX) Medium Deep Catalogues data. Funding for the SDSS and SDSS-II was provided by the Alfred P. Sloan Foundation, the Participating Institutions, the National Science Foundation, the U.S. Department of Energy, the National Aeronautics and Space Administration, the Japanese Monbukagakusho, the Max Planck Society, and the Higher Education Funding Council for England. The SDSS was managed by the Astrophysical Research Consortium for the Participating Institutions. GALEX (Galaxy Evolution Explorer) is a NASA Small Explorer, launched in 2003 April. We gratefully acknowledge NASA’s support for construction, operation, and science analysis for the GALEX mission, developed in cooperation with the Centre National d’Etudes Spatiales of France and the Korean Ministry of Science and Technology. This study also made use of Astropy, a community-developed core Python package for Astronomy (Astropy Collaboration, 2013, 2018). 
\end{acknowledgements}
\bibliography{aanda_p1}

\onecolumn
\appendix
\section{Structural parameters of the control sample of unbarred galaxies.}
\begin{table}[h!]
\captionsetup{width=.88\linewidth}  
\caption{The table contains information for the control sample of unbarred galaxies taken from NA10 and parameters obtained from GALFIT tool.}
\label{table:4}
\centering
\resizebox{0.88\textwidth}{!}{ 
\begin{tabular}{lllllllll}
\hline\hline\\
Galaxy & \multicolumn{1}{c}{\begin{tabular}[c]{@{}c@{}}RA\\ \\ (deg)\end{tabular}} & \begin{tabular}[c]{@{}l@{}}DEC\\ \\ (deg)\end{tabular} & \begin{tabular}[c]{@{}l@{}}Redshift\\ \\      z\end{tabular} & \begin{tabular}[c]{@{}l@{}}Stellar Mass\\ \\ $10^{10}M_{*}$\end{tabular} & \begin{tabular}[c]{@{}l@{}}HI Mass\\ \\ $10^{10}M_{*}$\end{tabular} & \multicolumn{1}{c}{\begin{tabular}[c]{@{}c@{}}r$_{bulge}$\\ \\ {[}mag{]}\end{tabular}} & \begin{tabular}[c]{@{}l@{}}r$_{e,bulge}$\\ \\ {[}kpc{]}\end{tabular} & n$_{bulge}$ \\ \\ \hline \\
1 & 40.06717  & -8.776953 & 0.0243 & 10.948 & -     & 15.83 & 1.16 & 2.66 \\ \vspace{0.1cm} 
2 & 315.28226 & -0.195178 & 0.0237 & 11.369 & -     & 15.66 & 0.96 & 2.03 \\ \vspace{0.1cm} 
3 & 352.60696 & 0.1567    & 0.0174 & 10.061 & 9.39  & 18.6  & 0.47   & 1.13 \\ \vspace{0.1cm} 
4 & 336.84125 & -1.161744 & 0.0592 & 11.244 & -     & 17.48 &    1.66     & 3.37     \\ \vspace{0.1cm} 
5 & 41.18923  & -8.164162 & 0.03   & 11.015 & -     & 16.2  & 0.80 & 2.1  \\ \vspace{0.1cm} 
6 & 205.5626  & 1.8574493 & 0.0286 & 11.163 & -     & 16.4  & 0.60 & 0.99 \\ \vspace{0.1cm} 
7 & 138.97916 & 10.132524 & 0.0313 & 11.07  & 10.37 & 15.4  & 1.75 & 1.81 \\ \vspace{0.1cm} 
8 & 221.06142 & 4.21855   & 0.0256 & 10.831 & 10.03 & 15.16 & 1.07  & 3.32 \\ \hline

\end{tabular}%
}
\end{table}

\end{document}